\documentclass[floatfix,notitlepage,aps,prd,twocolumn,superscriptaddress,showpacs,altaffilletter,preprintnumbers]{revtex4-1}
\usepackage{hyperref}
\hypersetup{
    colorlinks=true,
    linkcolor=blue,
}

\usepackage{graphicx}
\graphicspath{{img/}}

\newcommand{\plusminus}[2]{^{+#1}_{-#2}}

\usepackage{hepunits}

\usepackage{color}

\newcommand{\MkTable}[4]{
    \begin{table}[h!]
        \begin{tabular}{#2}#3\end{tabular}
        \caption{#4}
        \label{tbl:#1}
    \end{table}
}

\newcommand{\dtwoo}{$\text{D}_{2}\text{O}$}
\newcommand{\htwoo}{$\text{H}_{2}\text{O}$}
\newcommand{\pp}[1]{\left(#1\right)}
\newcommand{\textsub}[2]{#1_{\text{#2}}}
\newcommand{\sumon}[1]{\sum_{#1}}
\newcommand{\Nnmu}{N_{n}^{\pp{\mu}}}
\newcommand{\Nfmu}{N_{f}^{\pp{\mu}}}
\newcommand{\Nbkgmu}{N_{\text{bkg}}^{\pp{\mu}}}
\newcommand{\Nextmu}{N_{\text{ext}}^{\pp{\mu}}}
\newcommand{\Ncoincmu}{N_{\text{coinc}}^{\pp{\mu}}}
\newcommand{\Nradiomu}{N_{\text{radio}}^{\pp{\mu}}}
\newcommand{\Nmu}{N_{\mu}}
\newcommand{\lmu}{\ell_{\mu}}
\newcommand{\lavg}{\textsub{\ell}{avg}}

\newcommand{\eps}{\varepsilon}

\newcommand{\ecap}{\textsub{\eps}{Cap}}
\newcommand{\eobs}{\textsub{\eps}{Obs}}
\newcommand{\PNC}{P_{\text{NC}}}
\newcommand{\PB}{P_{\text{B}}}
\newcommand{\fB}{f_{\text{B}}}

\newcommand{\reffig}[1]{Figure~\ref{fig:#1}}
\newcommand{\reftbl}[1]{Table~\ref{tbl:#1}}
\newcommand{\refsec}[1]{Section~\ref{sec:#1}}

\newcommand{\refeq}[1]{Eq.~\eqref{eqn:#1}}

\newcommand{\dx}[1]{\text{d}#1}
\newcommand{\tmax}{t_{\text{max}}}

\newcommand{\Eeff}{E_{\text{eff}}}

\newcommand{\yieldunits}{\centi\meter\squared\per\left(\gram\cdot\mu\right)}
\newcommand{\syst}{\text{(syst.)}}
\newcommand{\stat}{\text{(stat.)}}
\newcommand{\hwtryield}{7.28 \pm 0.09\;\stat ^{+1.59}_{-1.12}\;\syst}
\newcommand{\saltyield}{7.30 \pm 0.07\;\stat ^{+1.40}_{-1.02}\;\syst}

\begin{document}

%\linenumbers

%\preprint{}

%\title{Study of Muon-Induced Neutrons at the Sudbury Neutrino Observatory}
%\title{Study of Cosmogenic Neutron Production
%       at the
%       Sudbury Neutrino Observatory}
\title{Cosmogenic Neutron Production
       at the
       Sudbury Neutrino Observatory}

%\author{SNO Collaboration (will get full list from Hamish)}
%\affiliation{}


%Authorlist v.18.1 Neutron November 2018.
%
\newcommand{\dec}{Deceased}
\newcommand{\alta}{Department of Physics, University of 
Alberta, Edmonton, Alberta, T6G 2R3, Canada}
\newcommand{\chicago}{Department of Physics, University of 
Chicago, Chicago IL} %60637
\newcommand{\ubc}{Department of Physics and Astronomy, University of 
British Columbia, Vancouver, BC V6T 1Z1, Canada}
\newcommand{\bnl}{Chemistry Department, Brookhaven National 
Laboratory,  Upton, NY 11973-5000}
\newcommand{\carleton}{Ottawa-Carleton Institute for Physics, Department of Physics, Carleton University, Ottawa, Ontario K1S 5B6, Canada}
\newcommand{\carletona}{Department of Physics, Carleton University, Ottawa, Ontario, Canada}
\newcommand{\uog}{Physics Department, University of Guelph,  
Guelph, Ontario N1G 2W1, Canada}
\newcommand{\lu}{Department of Physics and Astronomy, Laurentian 
University, Sudbury, Ontario P3E 2C6, Canada}
\newcommand{\lbnl}{Institute for Nuclear and Particle Astrophysics and 
Nuclear Science Division, Lawrence Berkeley National Laboratory, Berkeley, CA 94720-8153}
\newcommand{\lbla}{ Lawrence Berkeley National Laboratory, Berkeley, CA}
\newcommand{\lanl}{Los Alamos National Laboratory, Los Alamos, NM 87545}
\newcommand{\llnl}{Lawrence Livermore National Laboratory, Livermore, CA}
\newcommand{\lanla}{Los Alamos National Laboratory, Los Alamos, NM 87545}
\newcommand{\oxford}{Department of Physics, University of Oxford, 
Denys Wilkinson Building, Keble Road, Oxford OX1 3RH, UK}
\newcommand{\penn}{Department of Physics and Astronomy, University of 
Pennsylvania, Philadelphia, PA 19104-6396}
\newcommand{\pennx}{Department of Physics and Astronomy, University of 
Pennsylvania, Philadelphia, PA}
\newcommand{\queens}{Department of Physics, Queen's University, 
Kingston, Ontario K7L 3N6, Canada}
\newcommand{\uw}{Center for Experimental Nuclear Physics and Astrophysics, 
and Department of Physics, University of Washington, Seattle, WA 98195}
\newcommand{\uwx}{Center for Experimental Nuclear Physics and Astrophysics, 
and Department of Physics, University of Washington, Seattle, WA}
\newcommand{\uta}{Department of Physics, University of Texas at Austin, Austin, TX 78712-0264}
\newcommand{\triumf}{TRIUMF, 4004 Wesbrook Mall, Vancouver, BC V6T 2A3, Canada}
\newcommand{\ralimp}{Rutherford Appleton Laboratory, Chilton, Didcot, UK} %OX11 0QX
\newcommand{\iusb}{Department of Physics and Astronomy, Indiana University, South Bend, IN}
\newcommand{\fnal}{Fermilab, Batavia, IL}
\newcommand{\uo}{Department of Physics and Astronomy, University of Oregon, Eugene, OR}
\newcommand{\hu}{Department of Physics, Hiroshima University, Hiroshima, Japan}
\newcommand{\slac}{Stanford Linear Accelerator Center, Menlo Park, CA}
\newcommand{\mac}{Department of Physics, McMaster University, Hamilton, ON}
\newcommand{\doe}{US Department of Energy, Germantown, MD}
\newcommand{\lund}{Department of Physics, Lund University, Lund, Sweden}
\newcommand{\mpi}{Max-Planck-Institut for Nuclear Physics, Heidelberg, Germany}
\newcommand{\uom}{Ren\'{e} J.A. L\'{e}vesque Laboratory, Universit\'{e} de Montr\'{e}al, Montreal, PQ}
\newcommand{\cwru}{Department of Physics, Case Western Reserve University, Cleveland, OH}
\newcommand{\pnnl}{Pacific Northwest National Laboratory, Richland, WA}
\newcommand{\uc}{Department of Physics, University of Chicago, Chicago, IL}
\newcommand{\mitt}{Laboratory for Nuclear Science, Massachusetts Institute of Technology, Cambridge, MA 02139}
\newcommand{\ucsd}{Department of Physics, University of California at San Diego, La Jolla, CA }
\newcommand{	\lsu	}{Department of Physics and Astronomy, Louisiana State University, Baton Rouge, LA 70803}
\newcommand{\imp}{Imperial College, London, UK}%SW7 2AZ
\newcommand{\uci}{Department of Physics, University of California, Irvine, CA 92717}
\newcommand{\ucia}{Department of Physics, University of California, Irvine, CA}
\newcommand{\suss}{Department of Physics and Astronomy, University of Sussex, Brighton  BN1 9QH, UK}
\newcommand{\sussx}{Department of Physics and Astronomy, University of Sussex, Brighton, UK}
\newcommand{\lifep}{Laborat\'{o}rio de Instrumenta\c{c}\~{a}o e F\'{\i}sica Experimental de
Part\'{\i}culas, Av. Elias Garcia 14, 1$^{\circ}$, 1000-149 Lisboa, Portugal}
\newcommand{\lipx}{Laborat\'{o}rio de Instrumenta\c{c}\~{a}o e F\'{\i}sica Experimental de
Part\'{\i}culas,  Lisboa, Portugal}
\newcommand{\hku}{Department of Physics, The University of Hong Kong, Hong Kong.}
\newcommand{\aecl}{Atomic Energy of Canada, Limited, Chalk River Laboratories, Chalk River, ON K0J 1J0, Canada}
\newcommand{\nrc}{National Research Council of Canada, Ottawa, ON K1A 0R6, Canada}
\newcommand{\princeton}{Department of Physics, Princeton University, Princeton, NJ 08544}
\newcommand{\birkbeck}{Birkbeck College, University of London, Malet Road, London WC1E 7HX, UK}
\newcommand{\snoi}{SNOLAB, Lively, ON P3Y 1N2, Canada}
\newcommand{\snoix}{SNOLAB, Lively,  ON, Canada}
\newcommand{\uba}{University of Buenos Aires, Argentina}
\newcommand{\hvd}{Department of Physics, Harvard University, Cambridge, MA}
\newcommand{\pny}{Goldman Sachs, 85 Broad Street, New York, NY}
\newcommand{\pnv}{Remote Sensing Lab, PO Box 98521, Las Vegas, NV 89193}
\newcommand{\nts}{Nevada National Security Site, Las Vegas, NV}
\newcommand{\psis}{Paul Schiffer Institute, Villigen, Switzerland}
\newcommand{\liverpool}{Department of Physics, University of Liverpool, Liverpool, UK}
\newcommand{\uto}{Department of Physics, University of Toronto, Toronto, ON, Canada}
\newcommand{\uwisc}{Department of Physics, University of Wisconsin, Madison, WI}
\newcommand{\psu}{Department of Physics, Pennsylvania State University,
     University Park, PA}
\newcommand{\anl}{Deparment of Mathematics and Computer Science, Argonne
     National Laboratory, Lemont, IL}
\newcommand{\cornell}{Department of Physics, Cornell University, Ithaca, NY}
\newcommand{\tufts}{Department of Physics and Astronomy, Tufts University, Medford, MA}
\newcommand{\ucd}{Department of Physics, University of California, Davis, CA}
\newcommand{\unc}{Department of Physics, University of North Carolina, Chapel Hill, NC}
\newcommand{\dresden}{Institut f\"{u}r Kern- und Teilchenphysik, Technische Universit\"{a}t Dresden,  Dresden, Germany} %01069 
\newcommand{\isu}{Department of Physics, Idaho State University, Pocatello, ID}
\newcommand{\qmul}{School of Physics and Astronomy, Queen Mary University of London, UK}
\newcommand{\ucsb}{Dept. of Physics, University of California, Santa Barbara, CA}
\newcommand{\cern}{CERN, Geneva, Switzerland}
\newcommand{\utah}{Dept. of Physics, University of Utah, Salt Lake City, UT}
\newcommand{\casa}{Center for Astrophysics and Space Astronomy, University
of Colorado, Boulder, CO}%80309
\newcommand{\susel}{Sanford Underground Research Laboratory, Lead, SD}  %57754
\newcommand{\ntu}{Center of Cosmology and Particle Astrophysics, National Taiwan University, Taiwan}
\newcommand{\berlin}{Institute for Space Sciences, Freie Universit\"{a}t Berlin,
Leibniz-Institute of Freshwater Ecology and Inland Fisheries, Germany}
\newcommand{\bhsu}{Black Hills State University, Spearfish, SD} %57799-9003
\newcommand{\queensa}{Dept.\,of Physics, Queen's University, 
Kingston, Ontario, Canada} % K7L 3N6
\newcommand{\aasu}{Dept.\,of Chemistry and Physics, Armstrong  State University, Savannah, GA}
\newcommand{\ucb}{Physics Department, University of California at Berkeley, Berkeley, CA 94720-7300}
\newcommand{\ucbx}{Physics Department, University of California at Berkeley, and Lawrence Berkeley National Laboratory, Berkeley, CA}
\newcommand{\mcgill}{Physics Department, McGill University, Montreal, QC, Canada}% H3A 2T8
\newcommand{\columbia}{Columbia University, New York, NY}%10027
\newcommand{\rhul}{Dept. of Physics, Royal Holloway University of London, Egham, Surrey, UK}%  TW20 0EX
\newcommand{\ubama}{Department of Physics and Astronomy, University of Alabama, Tuscaloosa, AL}
\newcommand{\kit}{Institut f\"{u}r Experimentelle Kernphysik, Karlsruher Institut f\"{u}r Technologie, Karlsruhe, Germany}
\newcommand{\winnipeg}{Department of Physics, University of Winnipeg, Winnipeg, Manitoba, Canada}% 515 Portage Ave. R3B 2E9
\newcommand{\kwantlen}{Kwantlen Polytechnic University, Surrey, BC, Canada}% 12666 72nd Ave. V3W 2M8
\newcommand{\cea}{CEA-Saclay, DSM/IRFU/SPP, Gif-sur-Yvette, France}
\newcommand{\sunysb}{Laufer Center, Stony Brook University, Stony Brook, NY}% 11794
\newcommand{\rock}{Rock Creek Group, Washington, DC}
\newcommand{\rcnp}{Research Center for Nuclear Physics, Osaka, Japan}
\newcommand{\usd}{University of South Dakota, Vermillion, SD}
\newcommand{\lancaster}{Physics Department, Lancaster University, Lancaster, UK}%LA1 4YB,
\newcommand{\potsdam}{GFZ German Research Centre for Geosciences, Potsdam, Germany}%14473
\newcommand{\kirchhoff}{Ruprecht-Karls-Universit\"{a}t Heidelberg, Im Neuenheimer Feld 227, Heidelberg, Germany}%*D-69120
\newcommand{\continuum}{Continuum Analytics,  Austin, TX}% 221 W. 6th St. #1550,   78701
\newcommand{\gsu}{Dept. of Physics, Georgia Southern University, Statesboro, GA}
\newcommand{\pelmorex}{Pelmorex Corp., Oakville, ON} %2655 Bristol Circle, Oakville ON L6H7W1
\newcommand{\usaid}{Global Development Lab, U.S. Agency for International Development, Washington DC}%1300 Pennsylvania Ave NW, Washington, DC, 20004.
    
%%%%%%%%%%%%

%\affiliation{\aecl}
\affiliation{\alta}
\affiliation{\ucb}
\affiliation{\ubc}
\affiliation{\bnl}
%\affiliation{\uci}
\affiliation{\carleton}
%\affiliation{\chicago}
\affiliation{\uog}
\affiliation{\lu}
\affiliation{\lbnl}
\affiliation{\lifep}
\affiliation{\lanl}
\affiliation{\lsu}
\affiliation{\mitt}
%\affiliation{\nrc}
\affiliation{\oxford}
\affiliation{\penn}
%\affiliation{\princeton}
\affiliation{\queens}
%\affiliation{\ralimp}
\affiliation{\snoi}
\affiliation{\uta}
\affiliation{\triumf}
\affiliation{\uw}
\author{B.~Aharmim}\affiliation{\lu}
\author{S.\,N.~Ahmed}\affiliation{\queens}
\author{A.\,E.~Anthony}\altaffiliation{Present address: \usaid}\affiliation{\uta}
\author{N.~Barros}\altaffiliation{Present address: \pennx}\affiliation{\lifep}
\author{E.\,W.~Beier}\affiliation{\penn}
\author{A.~Bellerive}\affiliation{\carleton}
\author{B.~Beltran}\affiliation{\alta}
\author{M.~Bergevin}\altaffiliation{Present address: \llnl}\affiliation{\lbnl}\affiliation{\uog}
\author{S.\,D.~Biller}\affiliation{\oxford}
\author{R.~Bonventre}\affiliation{\ucb}\affiliation{\lbnl}
% ...
			
\collaboration{SNO Collaboration}
\noaffiliation

\date{\today}

\begin{abstract}

    Neutrons produced in nuclear interactions initiated by cosmic-ray muons present an irreducible background to many rare-event searches, even in detectors located deep underground. Models for the production of these neutrons have been tested against previous experimental data, but the extrapolation to deeper sites is not well understood. Here we report results from an analysis of cosmogenically produced neutrons at the Sudbury Neutrino Observatory. A specific set of observables are presented, which can be used to benchmark the validity of GEANT4 physics models. In addition, the cosmogenic neutron yield, in units of $10^{-4}\;\yieldunits$, is measured to be $\hwtryield$ in pure heavy water and $\saltyield$ in NaCl-loaded heavy water. These results provide unique insights into this potential background source for experiments at SNOLAB.

\end{abstract}

%\keywords{}

\maketitle

\section{Introduction}
\label{sec:intro}

    High energy muons created in cosmic-ray interactions in the Earth's atmosphere penetrate  deep underground, where they induce electromagnetic and hadronic showers.  These produce, among other particles of interest, free neutrons with an energy spectrum spanning several GeV. These cosmogenic neutrons form a direct background to searches for rare processes, such as neutrinoless double beta decay, nucleon decay, and dark matter interactions.

    The development and realization of next-generation detectors targeting these physics topics require unprecedented levels of background reduction. The prerequisite deep-underground location of such experiments reduces the rate of spallation backgrounds, but even the small number of remaining events can prove limiting to the potential physics reach of the experiments. It thus becomes critical to advance the understanding of the production and properties of cosmogenic neutrons. The average energy of the surviving cosmic muons increases with depth, and the extrapolation of cosmogenic neutron production rates from measurements made at shallow sites to greater depths is not well understood. Measurements at deep locations are critical to the success of future experiments.

    Many experimental collaborations have performed dedicated studies of cosmogenic neutrons using liquid targets~\cite{Soviet1973, Soviet1987, LSD, LVD, HertenbergerChen, CERNSPS, PaloVerde, Boulby, KamLAND, Borexino, KluckThesis, Zeplin, Aberdeen, DayaBay}, generally at relatively shallow depths. The deepest dedicated study to date was performed on data taken with the LSD detector, which was filled with liquid scintillator and located at a depth of 5200 meters water equivalent (m.w.e.)~\cite{LSD}.

    The Sudbury Neutrino Observatory (SNO) experiment offers a unique data set to study cosmogenic neutron production deep underground. The SNO detector was a kiloton-scale heavy water detector, located at a depth of $5890 \pm 94$~m.w.e. Using the parameterization found in~\cite{MeiHime}, the average muon energy at this depth is $\pp{363.0 \pm 1.2}$~GeV , higher than those in many other published studies~\cite{Soviet1973, Soviet1987, LVD, HertenbergerChen, CERNSPS, PaloVerde, Boulby, KamLAND, Borexino, KluckThesis, Zeplin, Aberdeen, DayaBay}, and comparable to that at LSD~\cite{LSD}. The SNO data can thus provide information in the high-energy regime, and further the understanding of how models for neutron production scale with muon energy.

    Here we present results derived from the observation of cosmogenic neutrons in the SNO detector, namely a comparison of observables to model predictions and a measurement of the neutron production rate. \refsec{detector} describes the SNO detector; \refsec{simulation} describes the Monte Carlo simulation used; \refsec{analysis} describes the analysis methods, including the selection criteria for muons and neutrons, and backgrounds to this measurement; \refsec{comparison} presents comparisons of characteristic observables seen in the data to those predicted by simulations; and \refsec{yieldresults} presents the results of the cosmogenic neutron yield measurement.

\section{The SNO Detector}
\label{sec:detector}

\begin{figure}
\includegraphics[width=\columnwidth]{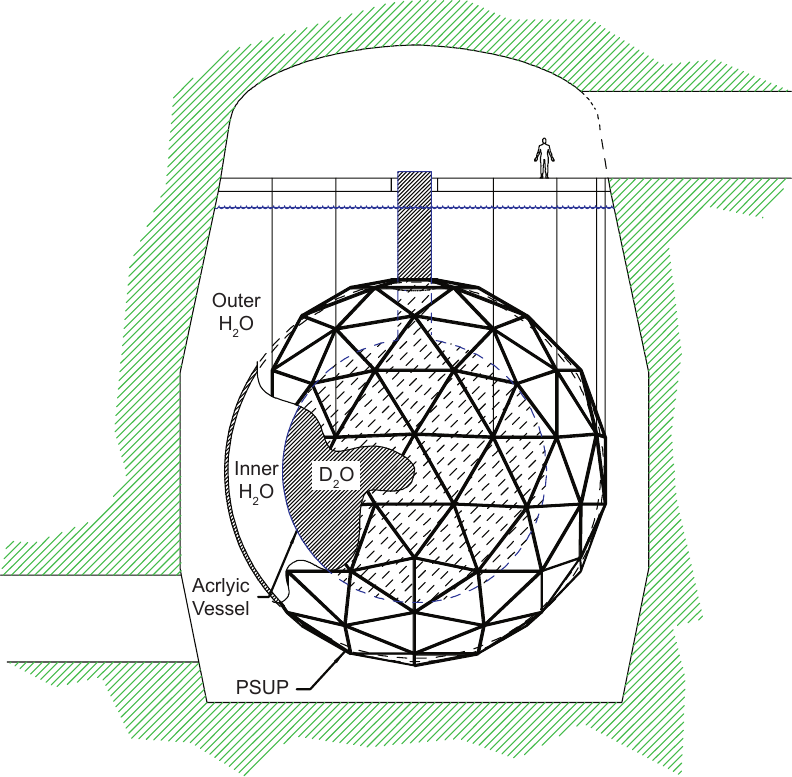}
\caption{\label{fig:sno-sketch}(Color online) Schematic diagram of the SNO detector.}
\end{figure}

    The SNO detector was a water Cherenkov detector located in INCO's (now Vale's) Creighton mine, near Sudbury, Ontario, at a depth of $\pp{2.092 \pm 0.033}$~km. It consisted of a spherical acrylic vessel (AV) 12~m in diameter, filled with 1000 metric tons of $99.92\%$ isotopically pure heavy water ($^{2}$H$_{2}$O, or \dtwoo{}). Surrounding the AV were 9456 Hamamatsu R1408 photomultiplier tubes (PMTs), each 20~cm in diameter, arranged onto a support structure (PSUP) of diameter 17.8~m. Each PMT was outfitted with a light-concentrator which increased the total photocathode coverage to approximately 55\%. The AV was surrounded by 7.4~kt of ultra-pure \htwoo{}. The detector arrangement is shown in \reffig{sno-sketch}.

    Data taking proceeded in three phases. During Phase I, the inner volume was filled with pure \dtwoo{}, with the neutron detection signal being the emission of a 6.25-MeV gamma following radiative capture on the deuteron. In Phase II, neutron detection was enhanced with the addition of 2~t of NaCl; $^{35}$Cl has a larger neutron capture cross section, and a cascade of photons totaling 8.6~MeV in energy is emitted upon neutron capture. In Phase III, an array of $^{3}$He proportional counters was deployed for neutron detection. The present analysis considers only data taken during the first two phases, with livetimes of $337.25 \pm 0.02$ and $499.45 \pm 0.02$ days, respectively.

\section{Monte Carlo Simulation}
\label{sec:simulation}

%\subsection{Overview}
\label{sec:mcintro}

The existing SNOMAN Monte Carlo and analysis code~\cite{SNONIM} incorporates a detailed, high-precision model of the SNO detector, including geometry, material and optical properties, and the response of the PMTs and electronic readout system.  This model was  based on measurements of microphysical parameters, and tuned and verified using calibration data from deployed radioactive and optical sources in the context of previous neutrino analyses~\cite{SNO2004,SaltPaper,SNOhep,SNOPhaseI,NCDPaper,LETA,SNOPhaseIII,SNOCombined,MuonAtmPaper,SNOnnbar}. However, the code relevant to the production and propagation of muons and neutrons evolved to become a compilation of algorithms from various sources. In particular, neutron propagation was based principally on the MCNP package~\cite{MCNP}, which in SNOMAN is applicable only for neutron energies below 20~MeV. For the purposes of both improved accuracy in the high-energy regime, and ease of interpretation by the scientific community, in the present analysis SNOMAN is used only for the purposes of modeling detector response and event reconstruction in the context of measuring the neutron yield; the propagation of muons and neutrons is performed in GEANT4~\cite{GEANT4} (version 10.00.p02), using the standard ``Shielding'' physics list with two modifications described below.

\label{sec:mcissues}

    In the course of this analysis, two issues concerning the treatment of deuterons by the standard physics processes included in the Shielding list were discovered. One of the most prominent neutron-producing reactions relevant to this analysis is the photonuclear reaction
    $\gamma d \rightarrow p n$,
    which can occur in electromagnetic showers initiated by a cosmic muon. GEANT4 tabulates photonuclear cross sections as a function of the mass number of the nucleus, but, when calculating the cross section for a given isotope, uses a mass number corresponding to the average mass of the naturally occuring isotopes of the given element. For heavy isotopes of hydrogen, this incorrectly returns the cross section on a free proton, which for energies below the pion threshold is 0, as no nuclear break-up can occur for a single nucleon. This issue has been reported to the GEANT4 development team and has been corrected in release version 10.5. In this work, a patch was implemented to disable this behavior for deuterons, for which a cross section tabulation already exists.

    It was further discovered that the default model for photonuclear final state generation, the Bertini Intranuclear Cascade, fails to properly model photodisintegration of the deuteron below the pion threshold. Indeed, while $\gamma d \rightarrow \gamma \gamma d$ and similar reactions occur, $\gamma d \rightarrow p n$ reactions do not. For the present analysis, we reimplemented the deuteron photodisintegration model developed for SNOMAN~\cite{LyonThesis} as a GEANT4 physics process, which is applied only to $\gamma d$ reactions below the pion threshold. In short, this model treats deuteron break-up as a two-body problem subject to conservation of energy-momentum. A summary of the contributions of various cosmogenic neutron-producing processes in GEANT4 is shown in \reftbl{productionprocesses}.

    \MkTable{productionprocesses}{l | c | c}{
        Process & Phase I & Phase II \\
        \hline
        Photonuclear        & 48.3\% & 46.1\% \\
        Neutron inelastic   & 25.1\% & 25.7\% \\
        $\pi$ inelastic     & 14.8\% & 16.1\% \\
        Proton inelastic    &  4.5\% &  4.7\% \\
        $\mu$ capture       &  3.3\% &  3.6\% \\
        $\mu$-nuclear       &  2.7\% &  2.4\% \\
        Other               &  1.3\% &  2.4\% \\
    }{Breakdown of cosmogenic neutron producing processes at SNO, as modeled by GEANT4. All processes labeled ``inelastic'' refer to inelastic scattering, and ``$\mu$-nuclear'' refers to direct muon-nucleus interactions via virtual photon exchange.}

%\subsection{Particle Generation}
\label{sec:mcinput}

    The first step in the Monte Carlo is to generate muons on a spherical shell approximately 4~m outside the PSUP. Given the spherical geometry of the SNO detector, the track can be specified using three coordinates: the impact parameter, which is the distance from the center of the detector to the midpoint of the line connecting the entrance and exit points; the zenith angle, which is the angle of the track measured from vertical; and the corresponding azimuthal angle. The impact parameters and entrance angles are sampled from the muons reconstructed in data, convolved with the resolution of the muon track reconstruction algorithm used in previous cosmic analyses~\cite{EMuS}. The initial muon energy is sampled from an analytic form taken from~\cite{MeiHime}, namely
\begin{equation}
    P\pp{E} = A e^{-bh\pp{\gamma - 1}}
                \pp{E + \eps\pp{1 - e^{-bh}}}^{-\gamma},
    \end{equation}
    where $b = 0.4/\text{km.w.e.}$, $\eps = 693$~GeV, $\gamma = 3.77$, are constants which parameterize the shape of the spectrum, $h = 5.89\text{ km.w.e.}/\cos\theta$ is the slant depth parameterized by the incident zenith angle $\theta$, and $A$ is the normalization. This distribution is the result of propagating muons from surface~\cite{GaisserBook}, neglecting their angular dependence, through a depth $h$, in the approximation of continuous energy loss. While the angular dependence of the energy spectrum at surface is neglected, the angular dependence due to the flat rock overburden is a larger effect, and is included.

    The propagation of muons and all daughter particles is handled by GEANT4, subject to the two corrections to photonuclear reactions described above. To mitigate poor performance due to the great number of low-energy photons created by high-energy muons, optical photon tracking is disabled and no detector response is simulated. All observables extracted from the Monte Carlo are thus taken as truth information, as output solely of the physics models.

\section{Analysis}
\label{sec:analysis}

    There are two goals of this study. The first is to provide a detailed comparison of the data to model predictions across a number of observables, including the capture time and the reconstructed position of the captured neutrons, offering validation of the models implemented in GEANT4. The second goal is a measurement of the neutron yield, defined as the number of neutrons produced per unit muon track length per unit target material, in the \dtwoo{} target.

    Use of a heavy water target in SNO offered a higher energy signature for neutron capture than the more-commonly used light water and liquid scintillator: neutron capture on the deuteron results in a $6.25$-MeV gamma, in comparison to the $2.2$-MeV gamma from capture on hydrogen. As a result the efficiency for detecting neutron capture events is greater than 95\% in the data set under consideration (Sec.~\ref{sec:obsefficiency}). The signal energy is also well above internal radioactive backgrounds, leading to effective neutron identification. Conversely, due to the relatively low muon flux at this depth, the data set is limited in statistics when comparing to studies performed of shallower sites.

\subsection{Muon reconstruction}

    The reconstruction of a muon candidate event is performed under the through-going hypothesis and outputs several parameters that specify the muon track, including the impact parameter ($b$) and zenith angle ($\theta$).
   
    The details of the reconstruction algorithm are described in~\cite{MuonAtmPaper}. The reconstruction is performed in two stages, where a preliminary fit from the first stage is used as the seed to a more sophisticated algorithm in the second stage. The first stage is a purely geometric construction: the entrance point is identified with the cluster of earliest hit PMTs, and the exit point with the charge-weighted position of all hit PMTs. The second stage, which takes this seed track as input, is a likelihood fit containing terms for the number of detected photoelectrons, and the PMT multiphotoelectron charge and hit-times. Using an external muon-tracking system to validate the fits, the muon reconstruction algorithm was found to perform with a resolution of less than 4~cm in impact parameter and $0.5\degree$ in zenith angle~\cite{EMuS}.

\subsection{Data selection}
\label{sec:dataset}

% {\it Does it work to have this after MC, since we simulate the muons in our data set?  That said, I like the idea of showing table 1 in this section.}
% Phases
% Event selection (muons and ns)
% Table of data / MC detected events (Tab 1 in unidoc)
    The data used in this analysis was taken during Phases I and II, with the AV filled with pure heavy water and salt-loaded heavy water, respectively. It is thus a subset of the data used in the SNO cosmic muon flux measurement~\cite{MuonAtmPaper}, which also considered data taken during Phase III, and the 13-day period between Phases II and III when the detector contained pure heavy water. Phase I data was collected between November 2, 1999 and May 28, 2001, and Phase II data was collected between July 26, 2001 and August 28, 2003, for a combined livetime of $836.7 \pm 0.03$ days.

    The selection criteria for muon events are designed to select through-going muons and reject instrumental backgrounds. Specifically, to qualify as a muon, events must have had at least 500 calibrated PMTs fired, with fewer than three of them in the neck of the AV, which is a characteristic of external light entering from the top of the detector. Events that occur within $5\;\micro\second$ of another event in which 250 PMTs fired, or within a $2$-s window containing 4 or more such events, are identified as a class of instrumental events called ``bursts,'' and are removed from the analysis. Furthermore, events with uncharacteristically low total PMT charge and/or broad timing distributions are inconsistent with the muon hypothesis, and are similarly identified as instrumental events. Further high-level cuts are made, among which are the requirements that the reconstructed impact parameter $b < 830$~cm to ensure the validity of the track fit, and the reconstructed energy loss $-\dx{E}/\dx{X} \geq 200$~MeV/m to reject muons that stop inside the detector volume. Finally, cuts are imposed on the fraction of photoelectrons geometrically contained inside the predicted Cherenkov cone for the muon track, and on the timing of these in-cone photons.

    These criteria are identical with previous cosmic muon analyses~\cite{MuonAtmPaper,EMuS} with one exception. A Fisher discriminant was previously used to reject stopping muons, but was found to incorrectly exclude muons with high light production --- potentially the most interesting from the standpoint of neutron production --- from the analysis. For the present analysis, we omit this linear discriminant cut; stopping muons are unlikely to contaminate neutron selection due to their relatively prompt decays, as discussed below. A total cross-sectional area of $216.4$~m$^{2}$ is considered in this analysis, for which Monte Carlo studies of cosmic muons in SNOMAN show the total event selection cut efficiency to be greater than 99\% for through-going muons~\cite{MuonAtmPaper}.

%   \redify{Interestingly, in the $\mu$nidoc this cut is attributed to rejecting both stopping muons {\it and} instrumentals... in the end it's effect is deemed ``negligible'' for cosmics, in agreement with what we see in applying the cuts. The language used in the muon paper is relevant to stopping muons, so that's what we say here. Any issues with accepting instrumentals/stopping $\mu$s here? Some info in $\mu$nidoc.}
   
    The average capture time for thermal neutrons is known to be on the order of tens of \milli\second{} in pure \dtwoo{}, and was decreased to a few \milli\second{} with the addition of NaCl in Phase II. We thus search for cosmogenic neutrons in a time window of $20\;\micro\second{} < \Delta t < \Delta\tmax{}$ following any through-going muon. The lower bound of $20\;\micro\second{}$ was chosen both to exclude Michel electrons from the decay of daughter muons from pions produced in hadronic showers, and to veto a period of several $\micro\second{}$ following particularly energetic muons in which the PMTs experienced significant afterpulsing. Imposing this lower bound reduces the livetime for neutron selection by less than 0.5\%. The upper bound $\Delta\tmax{}$ was chosen to accept $> 99\%$ of neutron captures in each phase, and is set to $300\;\milli\second{}$ in Phase I and $40\;\milli\second{}$ in Phase II. Low-level cuts to identify candidate events are identical to those used in previous analyses~\cite{NCDPaper, LETA, SNOCombined}. Neutron events are identified by reconstructing Compton scatters of the capture gammas under a single-scatter hypothesis, yielding a total effective electron energy $\Eeff$ and reconstructed radial position $r$. Neutron events are selected by requiring $4.0\;\MeV{} < \Eeff < 20.0\;\MeV{}$ and $r < 550.0\;\cm{}$.

    These high-level selection criteria differ from previous neutron selection in using a widened energy window consistent between the two phases, compared to the 6-10$\;\MeV{}$ window used previously for Phase II data~\cite{SaltPaper}, intended to maximize neutron acceptance.
   
    \reftbl{multiplicitycut} shows the number of muons accepted for the cosmogenic neutron search, and the percentages for which a follower was detected in both the data and Monte Carlo. The scarcity of neutron followers as shown in the table results in fewer than 3000 muons with detected neutron followers across both phases.

    \MkTable{multiplicitycut}{l | c | c  | c}{
        \- & \shortstack{\# Muons}
           & \shortstack{\% With followers \\ in data}
           & \shortstack{\% With followers \\ in MC} \\
        \hline
        Phase I & 21485
                      & $\pp{2.9 \pm 0.12}\%$
                      & $\pp{3.2 \pm 0.01}\%$ \\
        Phase II & 31898
                      & $\pp{5.8 \pm 0.13}\%$
                      & $\pp{5.7 \pm 0.01}\%$ \\
    }{The distribution of the number of muons included in this analysis, and fraction with followers, indicating the scarcity of neutron followers. The errors are statistical only.}

\subsection{Tests of model predictions}

    In order to validate the models of cosmogenic neutron production and propagation in the GEANT4 Shielding physics list at SNO depth and muon energies, we compare the data with model predictions for a number of observable distributions, including the properties of muons after which neutrons were observed, detected neutron multiplicity, neutron capture position, capture distance from the muon track, clustering of capture positions, and capture time.

    These quantities offer benchmarks of different aspects of the models implemented in GEANT4, and unique measurements of the physics involved in neutron production. For example, measurement of the per-muon neutron multiplicity yields insight into the validity of the cross sections of different neutron-producing reactions, while the capture time is sensitive to different neutron energies. Understanding these complementary observables in the simulations and the data will lead to improved physics modeling, imperative for more precise physics measurements.

    Furthermore, a measurement of the neutron production rate, using Monte Carlo information as input, requires the reliable simulation of several effects: direct and secondary production of neutrons, typically through electromagnetic and hadronic channels; the energy spectrum of produced neutrons, which can range up to several GeV; the transport of neutrons both at high and thermal energies; and the detection of capture gammas.

    As the neutrons are thermalized and then detected after radiative capture, this analysis is not directly sensitive to the energy of the neutrons, nor their production mechanisms. The observables listed above, however, allow a means to verify the reliability of the Monte Carlo implementations of neutron propagation and capture, in the context of measuring the neutron production rate.

\subsection{Neutron yield}
\label{sec:yieldoverview}

    The ``neutron yield'' is defined as the production rate of neutrons per unit muon track length per unit material density. Here we measure yields in heavy water, both pure and with the NaCl loaded at $0.2\%$ by weight. We define the track length of each muon as $\lmu = 2\sqrt{R_{AV}^{2} - b^{2}}$ through the target volume  of density $\rho$, where $R_{AV} = 600$~cm is radius of the AV, and $\Nnmu$ to be the number of neutrons produced by the muon. The yield is then
    \begin{equation}
    Y_{n} = \frac{1}{\rho}\frac{\sumon{\mu}\Nnmu}{\sumon{\mu}\lmu}
          = \frac{1}{\rho}\frac{\sumon{\mu}\Nnmu}{\Nmu\lavg}.
    \label{eqn:yield}
    \end{equation}
    where $\Nmu$ is the total number of muons and $\lavg$ is the average muon track length.

    In principle, the number of neutrons can be determined by simply counting neutron-like events following muons, with the following corrections: we express the probability for a neutron produced by a muon of impact parameter $b$ to be captured in the fiducial volume, the ``capture efficiency'', as $\ecap\pp{b}$; and the probability for a neutron capture at radius $r$ to trigger the detector and survive the event selection cuts, the ``observation efficiency'', as $\eobs\pp{r}$. With a background count of $\Nbkgmu$, the number of produced neutrons is then
    \begin{equation}
        N_{n}^{\pp{\mu}}
            = \frac{1}{\ecap\pp{b}}
                \pp{
                    \sumon{n = 1}^{\Nfmu}\frac{1}{\eobs\pp{r_{n}}} - \Nbkgmu
                },
    \label{eqn:neff}
    \end{equation}
    where $\Nfmu$ is the number of follower events, and we account for the relevant efficiencies on a per-neutron and per-muon basis. The number of background counts is
    \begin{equation}
        \Nbkgmu = \Nextmu + \Ncoincmu + \Nradiomu,
    \label{eqn:nbkg}
    \end{equation}
    comprised of neutrons originating external to the inner volume, radioactive backgrounds coincident with the follower selection window, and radioisotopic backgrounds also produced in spallation reactions, respectively. Estimates for the number of background counts in both phases are given in \refsec{backgrounds}.

    The first expression in \refeq{yield} is an idealized production rate, measured under the assumption that neutron production is a Poisson process, occurring constantly along the path of the muon. This is largely untrue, however, as the majority of production actually occurs during showering \cite{LiBeacomShowers}. The Poisson rate is equal to the mean per-muon yield were each muon to have equal track length. This is, in general, distinct from the mean of the true per-muon yield values calculated using the track length appropriate to each muon, which we denote by ${\bar Y}_{n}$. Because SNO is able to reliably reconstruct individual muon tracks, we calculate a per-muon yield
    \begin{equation}
    Y_{n}^{\pp{\mu}}
        = \frac{\Nnmu}{\rho\lmu}
    \label{eqn:simpleavg}
    \end{equation}
    unique to each muon, and compute ${\bar Y}_{n}$ as the mean $Y_{n}^{\pp{\mu}}$.

\subsection{Capture efficiency}
\label{sec:capefficiency}

\begin{figure}
    \includegraphics[width=\columnwidth]{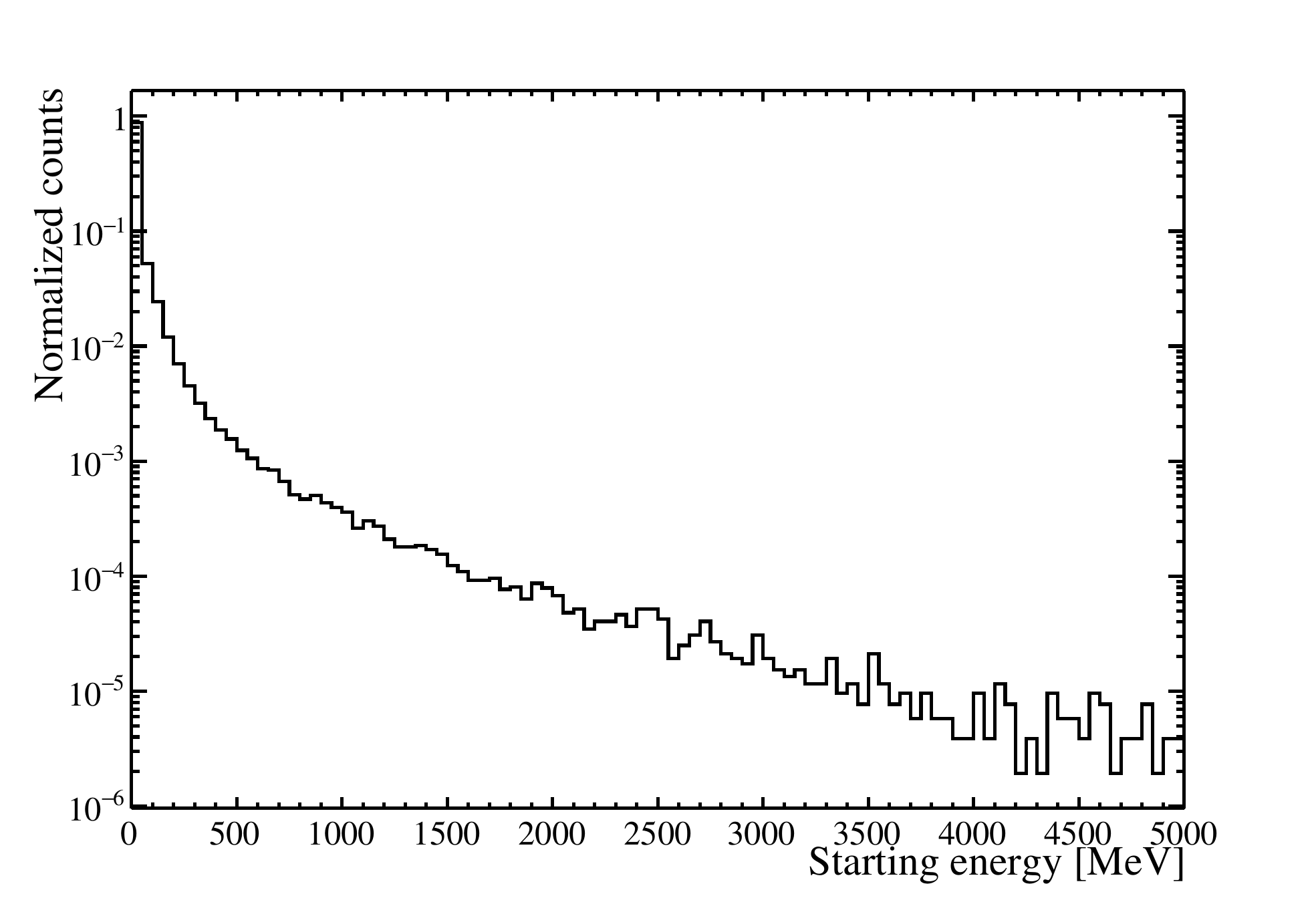}
    \caption{The spectrum of starting energies of muon-induced neutrons at SNO, as generated by GEANT4.}
    \label{fig:energyspectrum}
\end{figure}

\begin{figure}
    \includegraphics[width=\columnwidth]{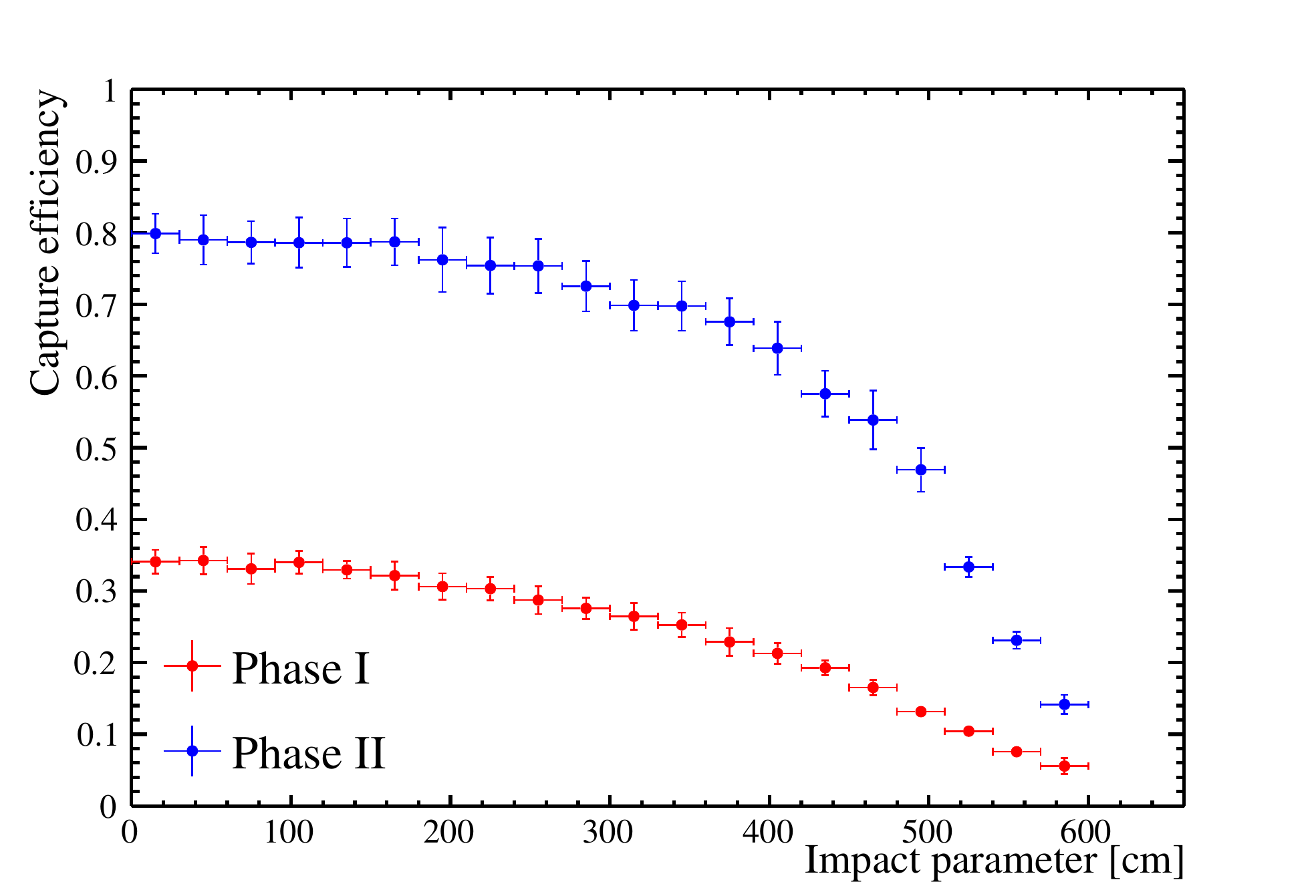}
    \caption{(Color online) GEANT4-based capture efficiencies for cosmogenic neutrons in Phases I (red) and II (blue). Error bars represent the spread in efficiency due to the neutron energy spectrum.}
    \label{fig:capeffcurves}
\end{figure}

    The capture efficiency is defined as the fraction of neutrons produced by a muon that are captured in the fiducial volume, parameterized as a function of the impact parameter of the muon. A $^{252}$Cf source was deployed in SNO to measure the capture efficiency of MeV-scale neutrons (see \reffig{cf252}), but the energy spectrum from cosmogenic production extends much higher, and the capture efficiency in this regime may be different. We thus evaluate this efficiency solely using GEANT4 simulations. An uncertainty on the capture efficiency due to the spectrum of starting neutron energies, shown in \reffig{energyspectrum}, is calculated by computing the efficiency in ten bins in energy, ranging from 0 to 5~GeV, and computing the RMS difference of these binned efficiencies from the nominal value, weighted by each bin's integral of the energy spectrum. The capture efficiencies in both phases are shown in \reffig{capeffcurves}. The cosmogenic capture efficiency curves differ from those measured with the $^{252}$Cf source (see \reffig{cf252}) for two reasons: principally, the cosmogenic capture efficiency is parameterized by the muon impact parameter, not neutron starting position, and also differences in the neutron energy spectra.

\subsection{Observation efficiency}
\label{sec:obsefficiency}

\begin{figure}
    \includegraphics[width=\columnwidth]{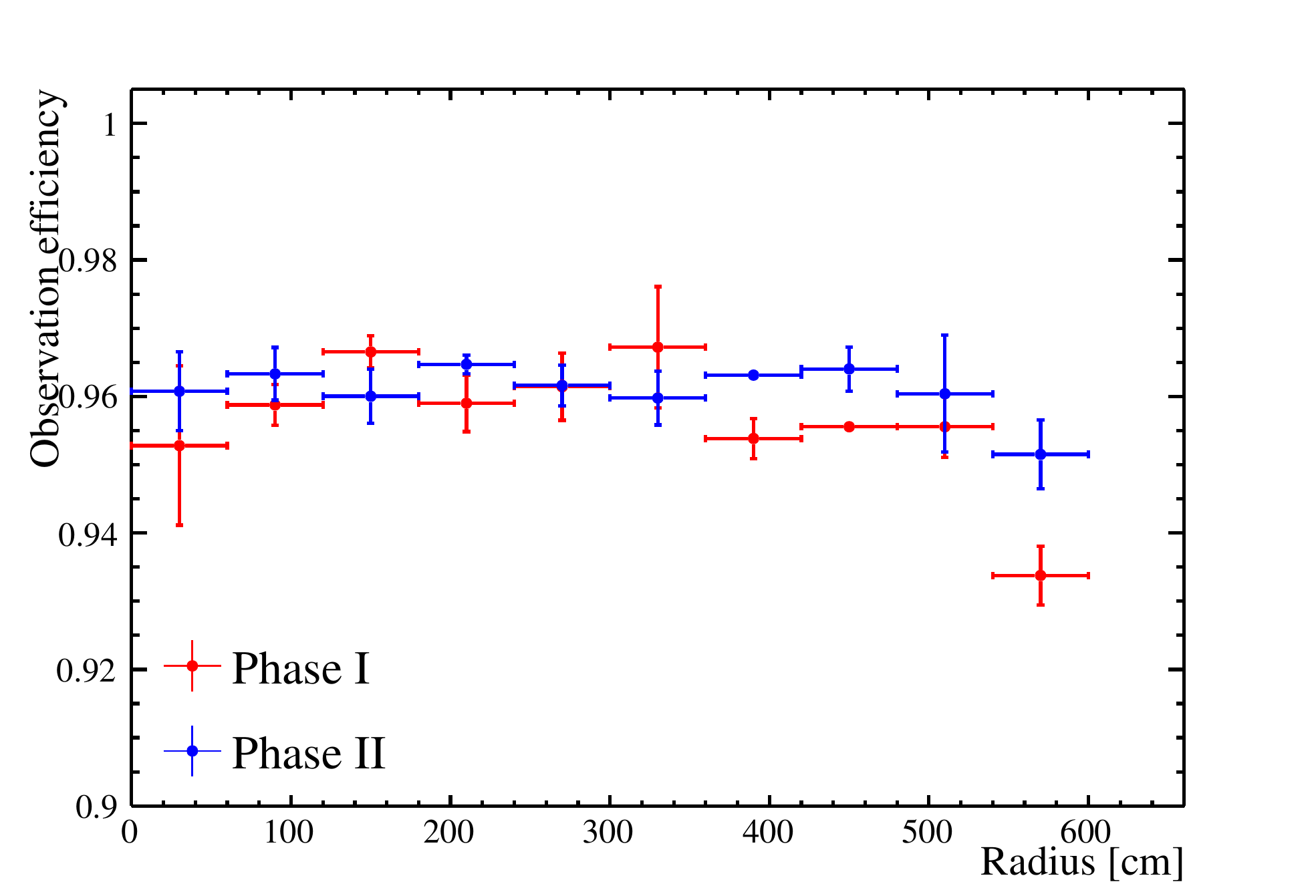}
    \caption{(Color online) SNOMAN-based observation efficiency for neutron captures on D in Phase I (red), and $^{35}$Cl in Phase II (blue). Error bars are statistical.}
    \label{fig:obseffcurves}
\end{figure}

    The observation efficiency is defined as the probability for a neutron capture through a visible capture mode to trigger the detector and pass the event selection criteria outlined in \refsec{dataset}. We evaluate this efficiency by propagating and reconstucting capture gammas in SNOMAN. This efficiency is shown in \reffig{obseffcurves}. Because the energy threshold used in this analysis is lower than that used in past solar neutrino analyses, this efficiency is comparable in both phases, and relatively stable with respect to position in the detector.

\subsection{Backgrounds}
\label{sec:backgrounds}

    The yield measurement as defined in Equations \eqref{eqn:yield} and \eqref{eqn:neff} is subject to three general classes of background, namely cosmogenic neutrons from sources other than the detector volume, radioisotopes produced in conjunction with neutrons, and random coincident events, each of which is discussed below.

\subsubsection{External captures}
\label{sec:externals}

    One background to measuring the rate of neutron production in heavy water is contamination from cosmogenic neutrons produced in other materials, which we define as ``external captures.'' At SNO, the principal external sources are the AV and surrounding light water. We assess this contamination as a function of impact parameter, and find, using GEANT4, that the average number of external neutrons capturing in the fiducial volume per muon is at most
    $(5.3 \pm 0.2) \times 10^{-3}$
    in Phase I and
    $(1.5 \pm 0.1) \times 10^{-2}$
    in Phase II, where the larger capture efficiency in Phase II determines the difference.

\subsubsection{Cosmogenic radioisotopes}
\label{sec:radioisotopes}

\begin{figure*}
    \includegraphics[width=\columnwidth]{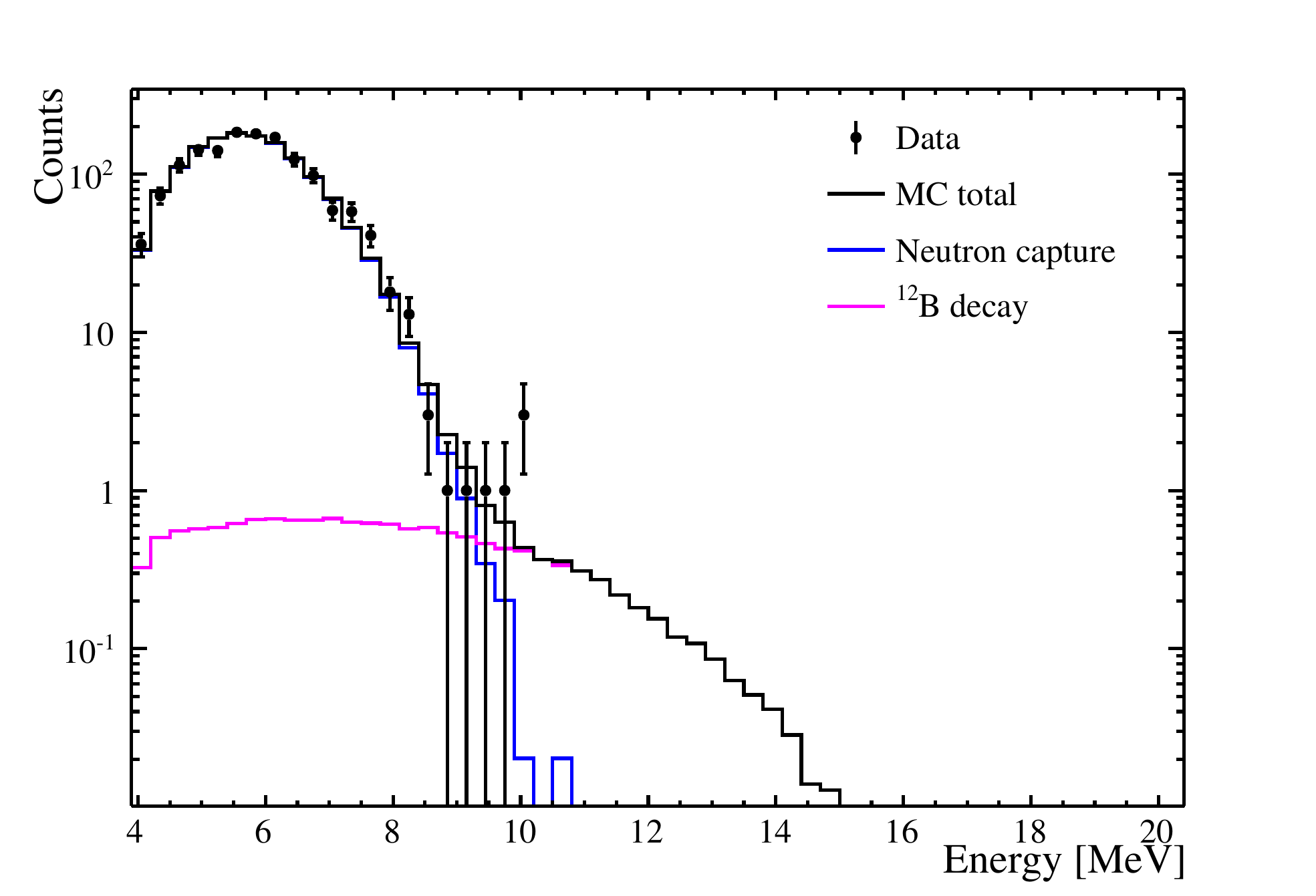}
    \includegraphics[width=\columnwidth]{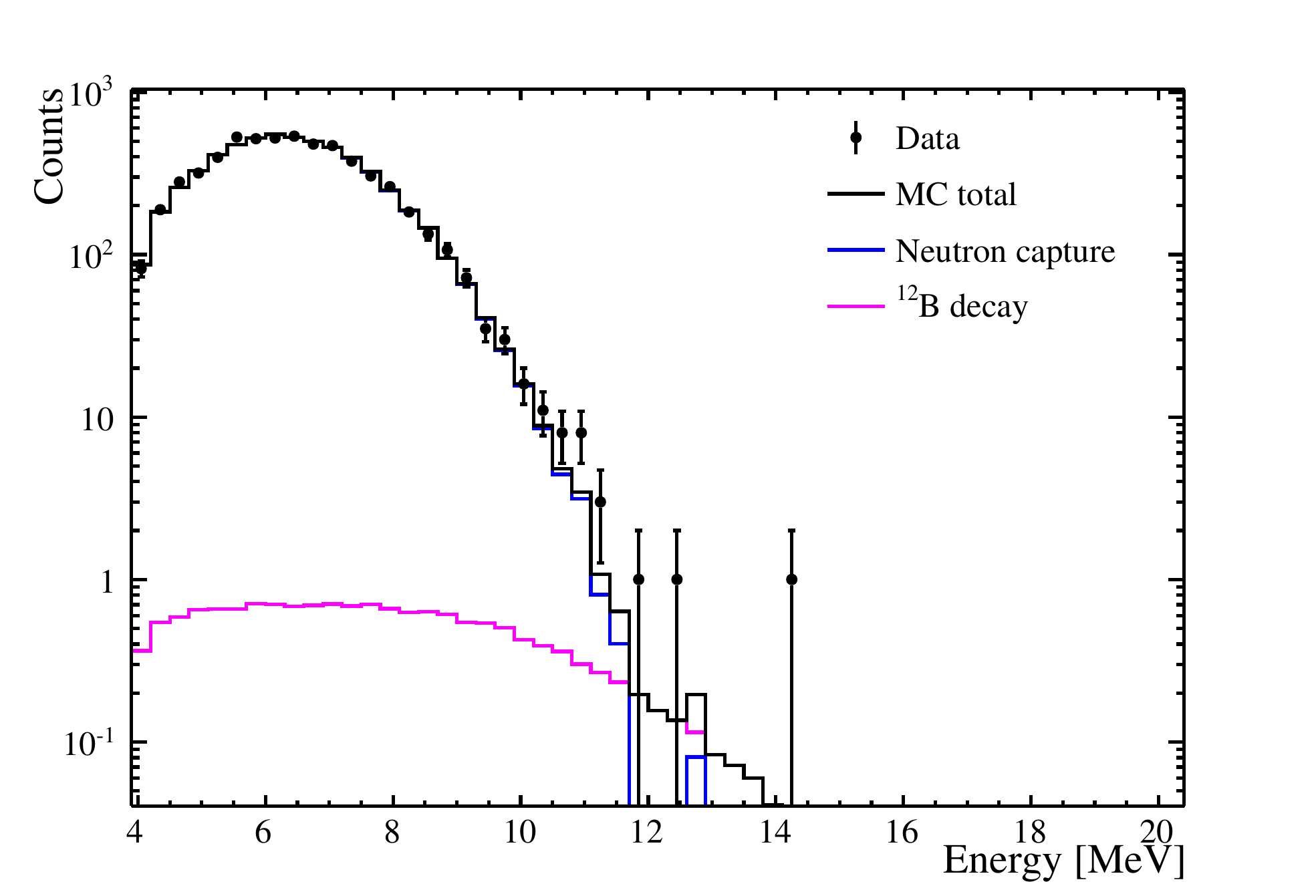}
    \caption{(Color online) Determination of $^{12}$B contamination, in Phases I (left) and II (right). The time delay and reconstructed energy (shown here) distributions are fit to a combination of exponentials, corresponding to neutron captures and $^{12}$B decays.}
    \label{fig:12Benergy}
\end{figure*}

    The passage of a muon can result in the production of various unstable isotopes~\cite{LiBeacomSpallation}, as well as the neutrons that are the focus of this analysis. While the usual concern for cosmogenic production centers on long-lived isotopes, such as $^{16}$N with a half-life of roughly $7\;\second{}$, the timing cut used to select followers makes this analysis sensitive to the production of short-lived isotopes. From both calculations and measurements of isotope production at Super-Kamiokande~\cite{LiBeacomSpallation, SuperK}, we determine the expected dominant isotope background to be $^{12}$B, a beta-emitter with a half-life of $20\;\milli\second{}$ and $Q$-value of $13\;\MeV{}$. Our approach to assessing the contribution of this background is data-driven: we search for contamination from $^{12}$B decays using a maximum likelihood fit of both the timing and energy distributions of events following cosmic muons. Explicitly, where $t$ and $E$ are the time delay and energy of each event, we construct a likelihood function
    \begin{equation}
        L\pp{\tau, \fB} = \prod\limits_{\text{events}}\pp{
               \frac{1 - \fB}{\tau}e^{-t/\tau}\PNC\pp{E}
             + \frac{\fB}{\tau_{1}}e^{-t/\tau_{1}}\PB\pp{E}
           }
    \end{equation}
    where $\tau_{1} = 20\;\milli\second{}/\ln{2}$ is the $^{12}$B lifetime, and $\PNC$ and $\PB$ are the reconstructed energy spectra for neutron captures and $^{12}$B $\beta$-decays, respectively. The fit parameters are $\tau$, the neutron capture time, and $\fB$, the fractional $^{12}$B contamination. The fit is performed separately on the samples of follower events in each phase; the results of the fit in energy space are shown in \reffig{12Benergy}. The best fit capture time constants are consistent with those fit under the boron-free hypothesis (\refsec{delay}).

    We compute an upper limit on the fractional $^{12}$B contamination at the 90\% confidence level by marginalizing over the free time constant. This results in limits on the radioisotopic contamination of $2.4\%$ and $0.67\%$ in Phases I and II, respectively, which are included as uncertainties on the measured neutron yield.

\subsubsection{Random coincidences}
\label{sec:coincidents}

    All remaining backgrounds are uncorrelated with the passage of a muon, and are classified as random coincidences. We assess this class of backgrounds by imposing neutron selection criteria on events in a $3$-s time window immediately preceding the trigger time of each muon. Doing so determines the average coincidence rates to be $7.89\times10^{-4}\;\second^{-1}$ and $9.73\times10^{-4}\;\second^{-1}$ in Phases I and II, respectively, which translate to average numbers of coincident events per muon of $2.4\times10^{-2}$ and $3.9\times10^{-3}$, respectively.

\section{Study of Event Distributions}
\label{sec:comparison}

    To aid in the development and improvement of physical models, both strictly theoretical and those implemented in simulation packages, we present distributions of observables of cosmogenic neutrons and their relation to their leading muon in the data, and a comparison to model-based predictions. Specifically, we show distributions of the track parameters of muons for which neutron followers were observed, follower multiplicity, the capture positions measured both in the detector and in relation to the leading muon, and the time delay between the muon and follower event. In all cases, the MC has been scaled to the normalization of the data, for easy comparison of the shapes of the distributions.

\subsection{Follower selection}
\label{sec:followerselection}

\begin{figure*}
    \includegraphics[width=\columnwidth]{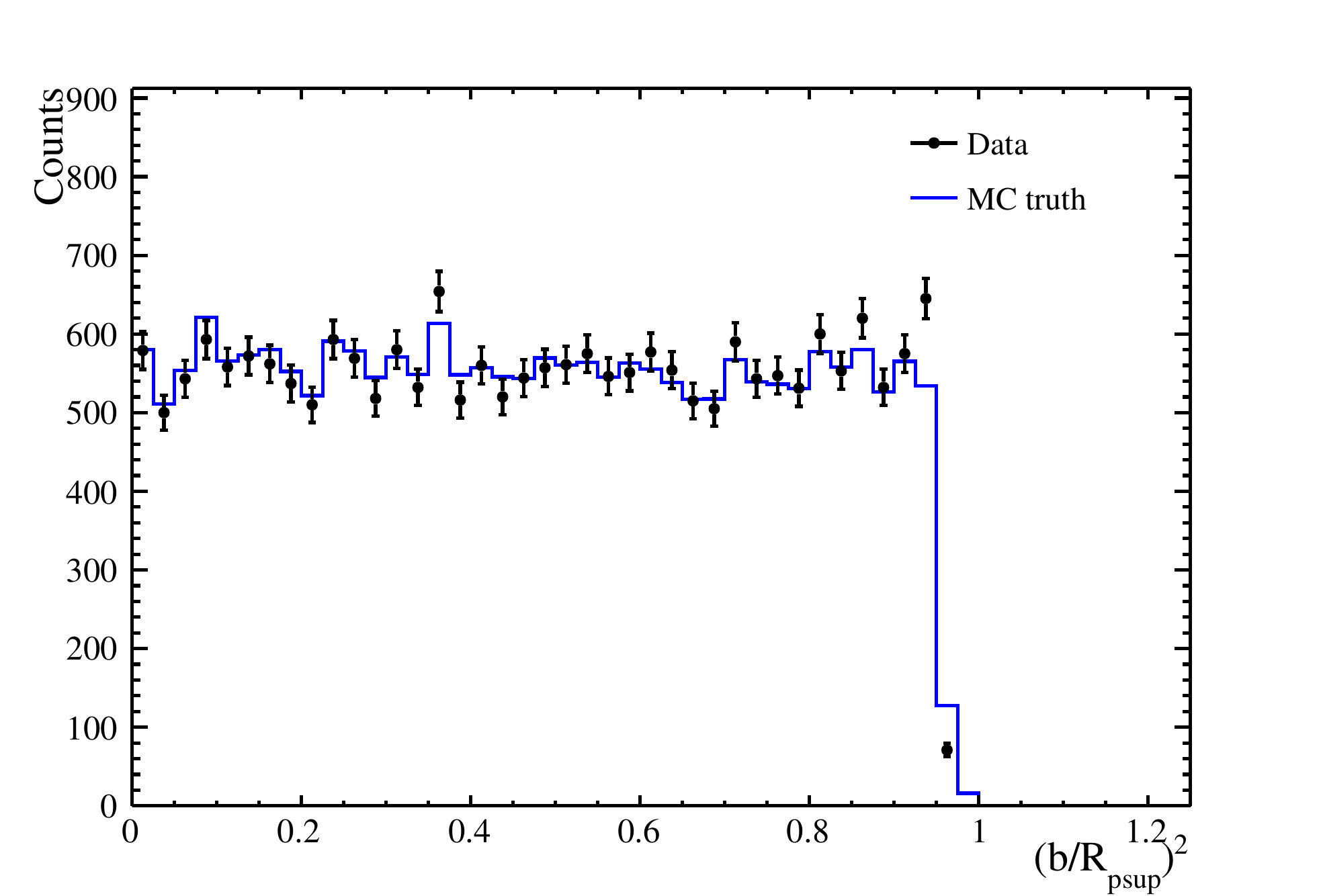}
    \includegraphics[width=\columnwidth]{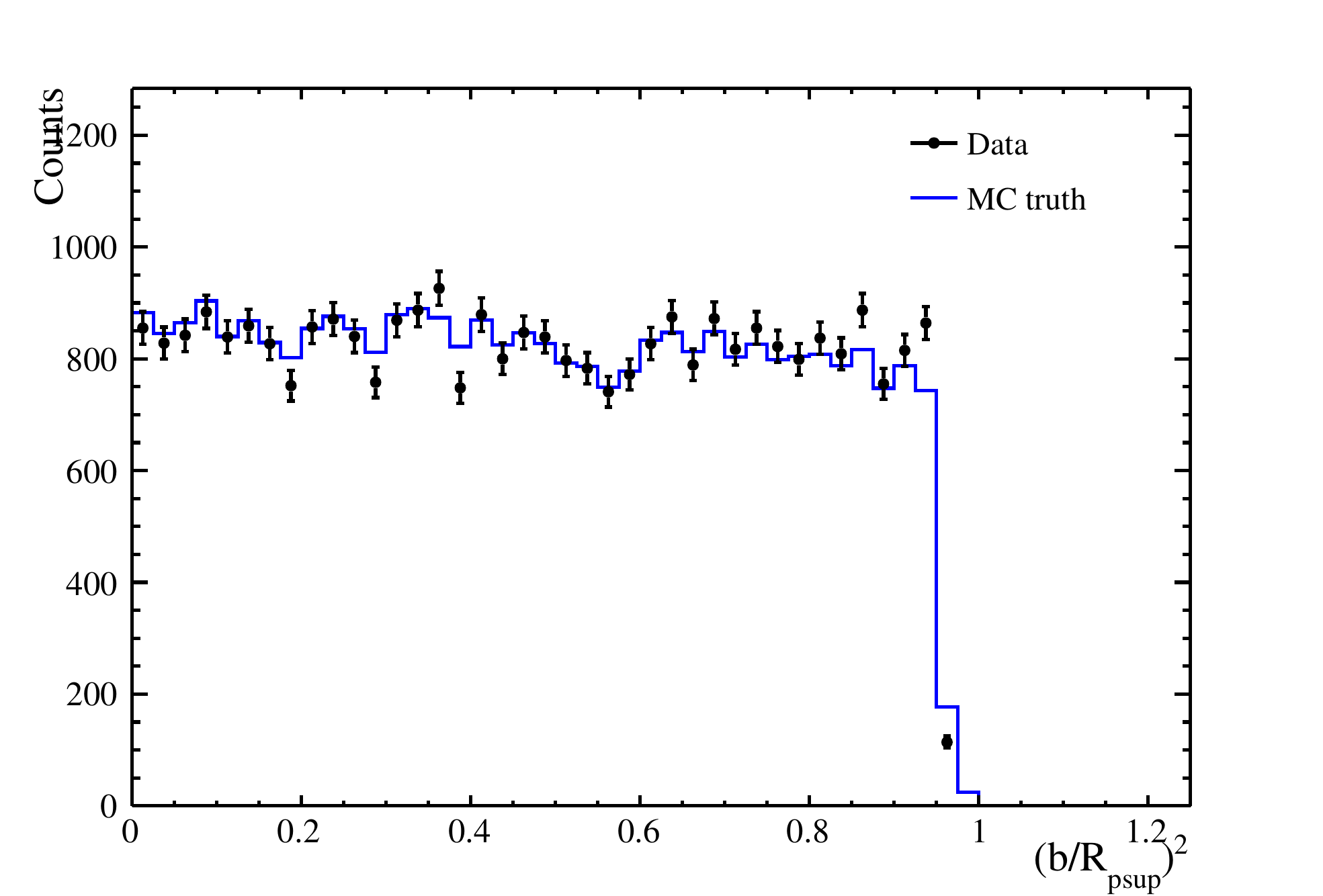}
    \includegraphics[width=\columnwidth]{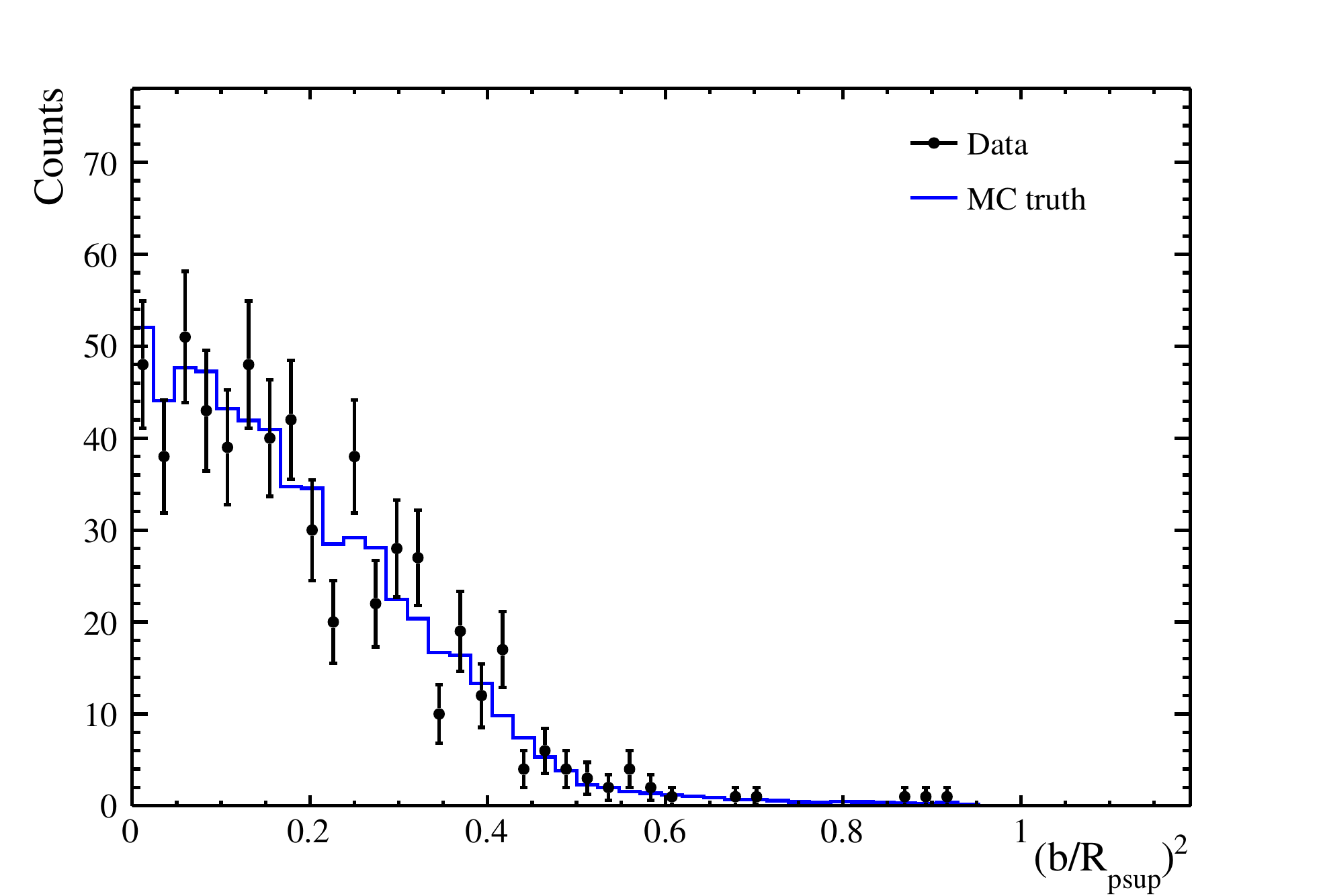}
    \includegraphics[width=\columnwidth]{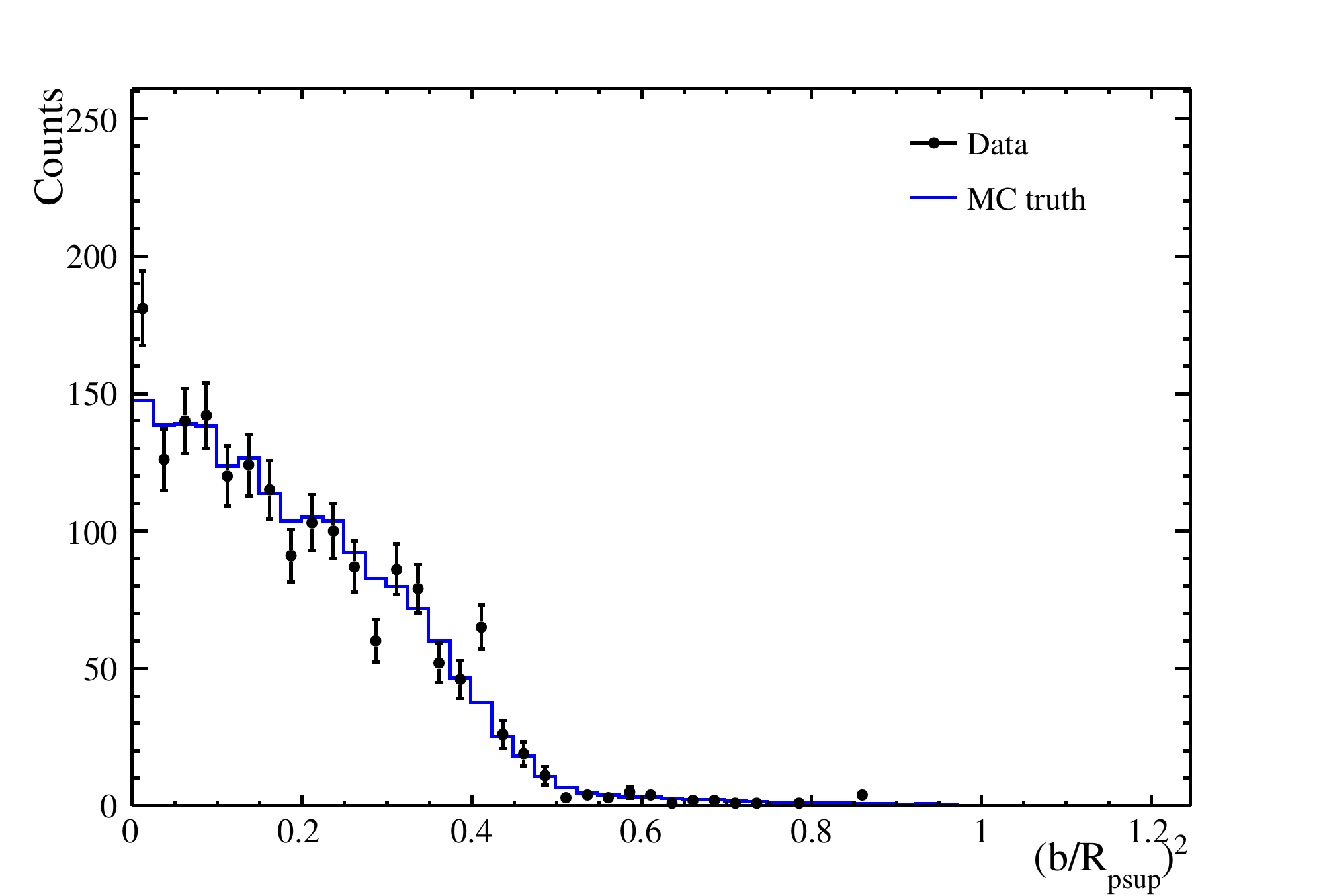}
    \caption{(Color online) Area-normalized impact parameters $b^{2}/R_{\text{PSUP}}^{2}$ of all muons (top) and only muons with followers (bottom), in Phase I (left) and II (right). $R_{\text{PSUP}} = 850$~cm is the radius of the PSUP. The AV boundary is at abscissa value $\approx 0.5$.}
    \label{fig:impactcomparison}
\end{figure*}

\begin{figure*}
    \includegraphics[width=\columnwidth]{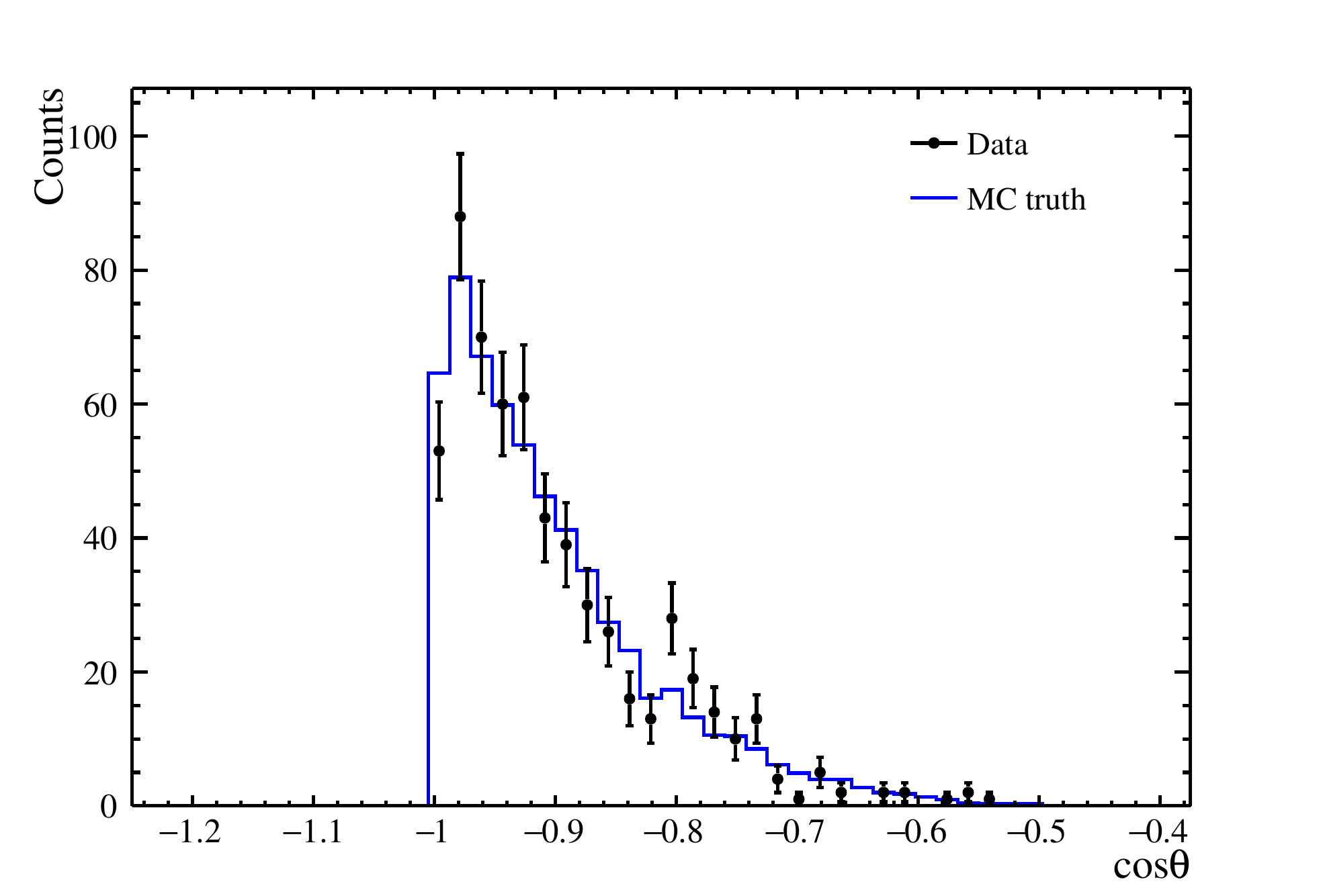}
    \includegraphics[width=\columnwidth]{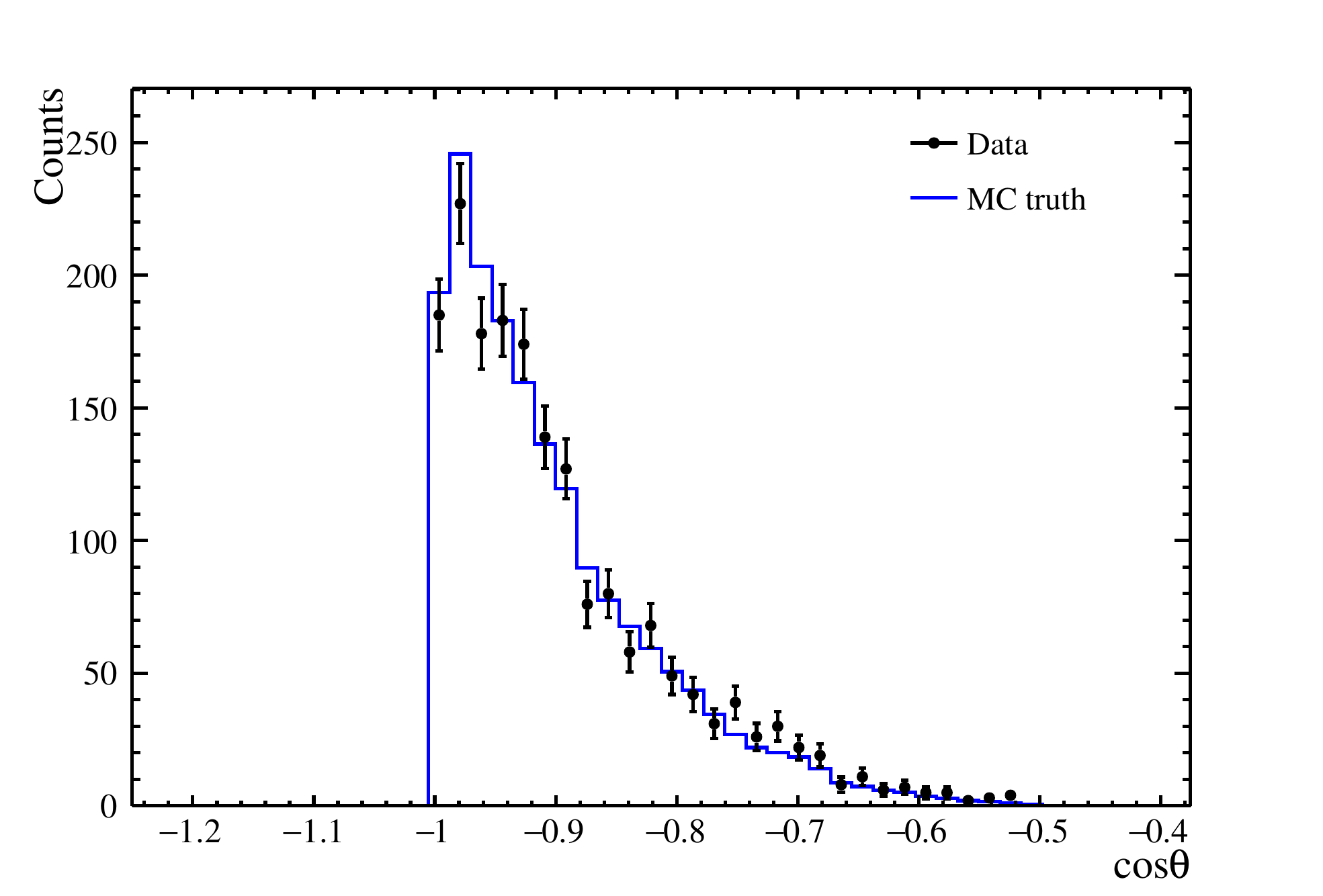}
    \caption{(Color online) Entrance zenith angles of muons with detected followers, in Phases I (left) and II (right).}
    \label{fig:cosqcomparison}
\end{figure*}

    The number of muons that have follower events passing the selection criteria described in \refsec{dataset} is shown in \reftbl{multiplicitycut}. The enhanced proportion of muons after which followers were observed in Phase II reflects the higher capture cross section. \reffig{impactcomparison} shows the distributions of muon impact parameter, both for all muons and only those with followers. The pre-selection distributions agree because the input to the Monte Carlo is sampled from the population of muons observed in the data. The shapes of the post-selection distributions are roughly proportional to the muon track length in the detector. With regard to the zenith angle, the subset of muons with followers is representative of the larger population, and is shown in \reffig{cosqcomparison}.

\subsection{Follower multiplicity}
\label{sec:multiplicity}

\begin{figure*}
    \includegraphics[width=\columnwidth]{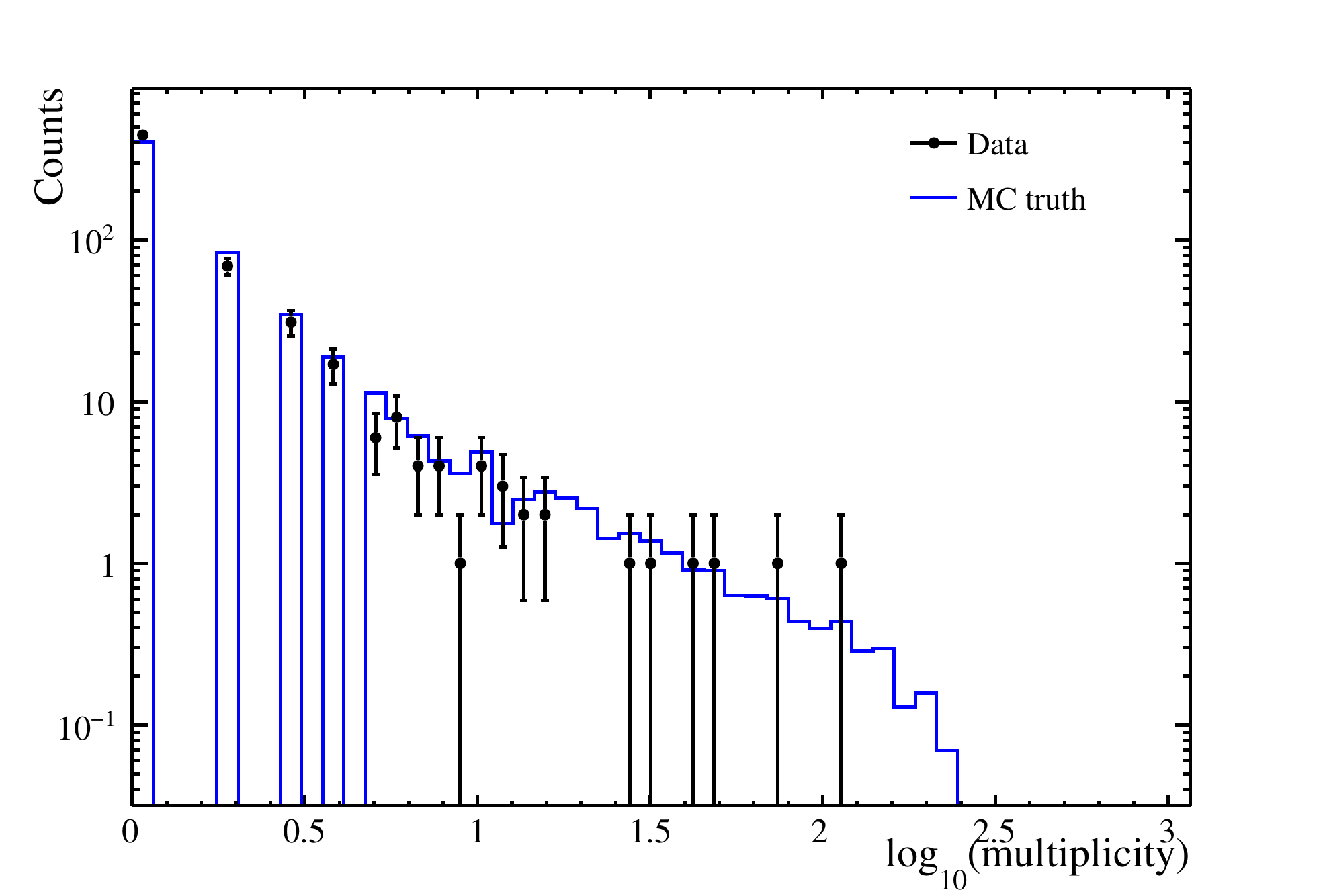}
    \includegraphics[width=\columnwidth]{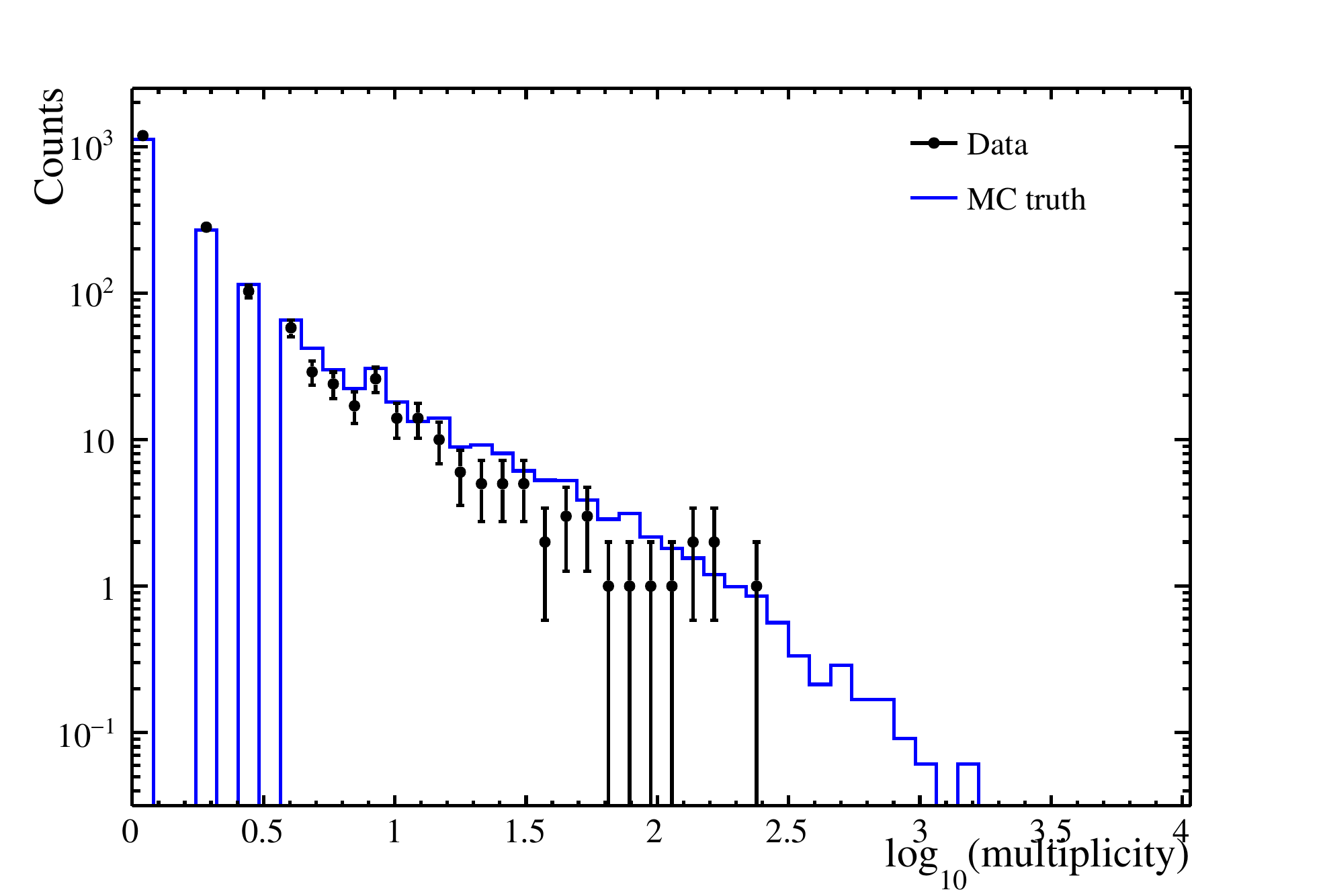}
    \caption{(Color online) Number of detected neutron followers per muon, in Phases I (left) and II (right). Each entry to the histograms represents one muon.}
    \label{fig:multiplicitycomparison}
\end{figure*}

    The distributions of the number of neutron-like events following a muon are shown in \reffig{multiplicitycomparison}. Muons with hundreds of followers were observed in each phase; indeed, events of such high multiplicity are reproduced using existing simulation tools. The potential disagreement in the number of high-multiplicity events in Phase II, however, may indicate that some reactions on chlorine are mismodeled. This could be attributed to incorrect cross sections for the dominant, low-multiplicity, neutron-producing processes, i.e. photonuclear and neutron inelastic scattering, or incorrect final-state generation after near-complete nuclear breakup at high energies.

    Distinct identification of cosmic muons as showering either electromagnetically or hadronically has been demonstrated by studying the distribution of multiplicities of neutron followers in high energy ($> 90\;\GeV{}$) muon-induced showers in liquid scintillator detectors~\cite{ShowerSeparation}. When imposing shower selection criteria, the multiplicity distribution analagous to those shown in \reffig{multiplicitycomparison} exhibited two peaks, corresponding to electromagnetic and hadronic showering, with the hadronic case corresponding to larger multiplicities. Our data set includes neutrons of all origins, and the distributions shown in \reffig{multiplicitycomparison} do not exhibit the bimodal topography characteristic of such shower separation.

\subsection{Capture position}
\label{sec:rcubed}

\begin{figure*}
    \includegraphics[width=\columnwidth]{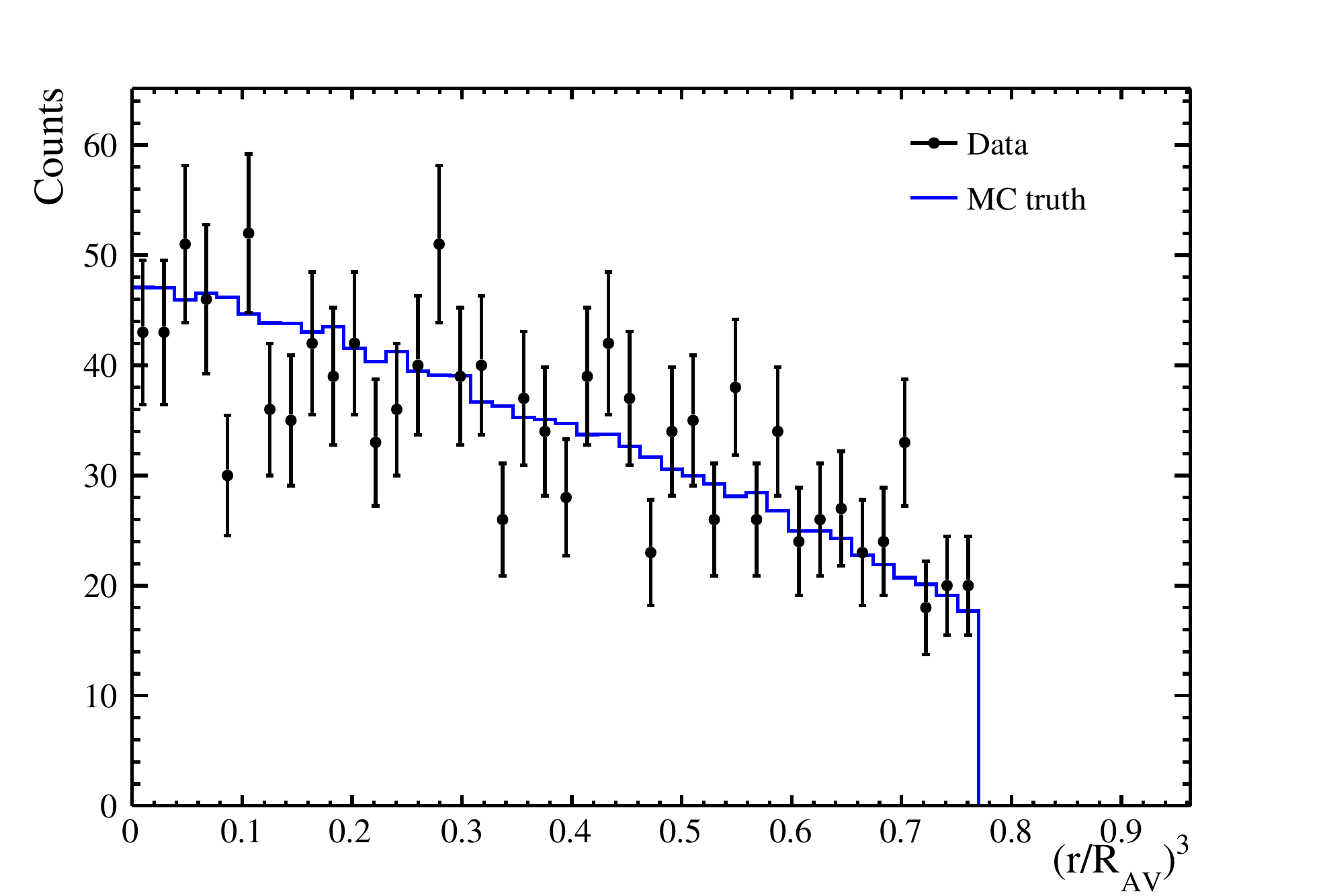}
    \includegraphics[width=\columnwidth]{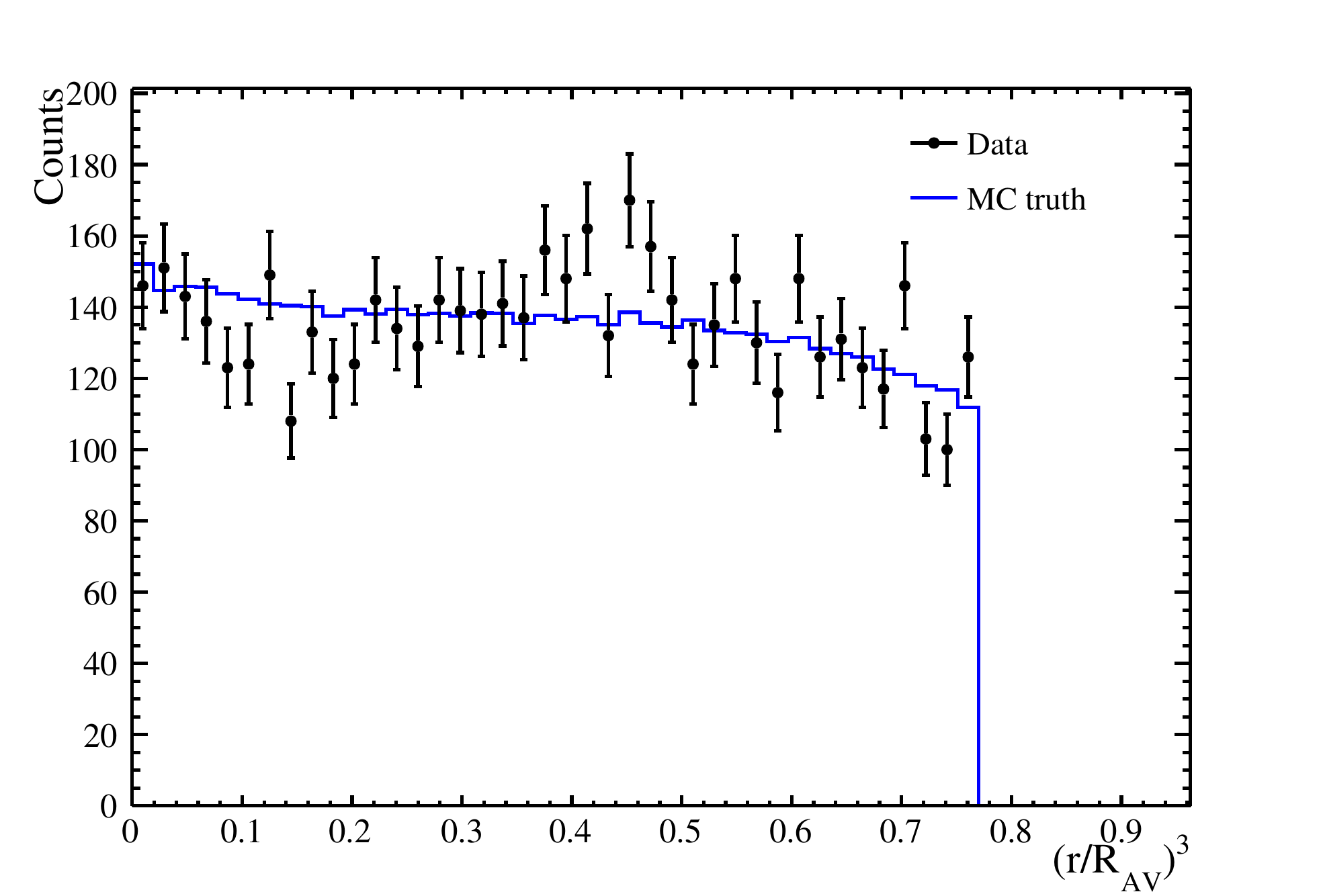}
    \caption{(Color online) Volume-normalized capture position $r^{3}/R_{\text{AV}}^{3}$ of detected followers, in Phases I (left) and II (right). $R_{\text{AV}} = 600$~cm is the radius of the AV.}
    \label{fig:rcubedcomparison}
\end{figure*}

    \reffig{rcubedcomparison} shows the distributions of the radial position of neutron captures in the detector. Because the muon flux is uniform in area and, in aggregate, neutrons are produced uniformly along a track, they are, in aggregate, produced uniformly in the volume of the detector. This is reflected in Phase II, where there is a large capture cross section and the capture position is more strongly correlated with production position. In Phase I, where the effective capture cross section is reduced by 2 orders of magnitude, neutrons are more likely to diffuse out of the fiducial volume; this effect grows as the muon and, hence, neutrons are located closer to the edge of the AV, which has a relatively high hydrogen content, and results in a deficiency of captures in the outer fiducial volume compared to the center. The agreement of the comparison shown in \reffig{rcubedcomparison} constitutes a partial validation of the propagation of neutrons in the GEANT4 detector model, but is complicated by the finite size of the detector. More ideal tests would use large volumes where boundary effects are suppressed.

\subsection{Capture clustering}
\label{sec:bunchiness}

    The majority of neutron production occurs in electromagnetic showers. The initiation of a shower usually entails a very localized energy deposition by the muon, in contrast to the smaller, constant ionization losses. In the electromagnetic case, this energy deposition has a characteristic profile in the direction of the muon track, which at cosmic-muon energies in light water has a width typically on the order of several meters; see~\cite{LiBeacomShowers} for a discussion.

    In an attempt to profile the energy deposition relevant to neutron production, we investigate the clustering of muon-induced neutrons. Specifically, we use the neutron capture positions as proxies for their production positions, which act as proxies for energy deposition. We define a clustering metric, $\textsub{\sigma}{Long}$, as the standard deviation in the coordinate of the followers' capture positions measured longitudinally along the muon track. More specifically, we define $\vec{r}_{n}$ as the reconstructed position of a neutron capture event, $\vec{r}_{\mu\text{ entrance}}$ and $\vec{r}_{\mu\text{ exit}}$ as the positions where the muon enters and exits the PSUP, respectively, and $x_{n}$ as the coordinate of the neutron capture measured along the track; that is,
    \begin{equation}
     x_{n}
        = \frac{
            \pp{\vec{r}_{n} - \vec{r}_{\mu\text{ entrance}}} \cdot
                \pp{\vec{r}_{\mu\text{ exit}} - \vec{r}_{\mu\text{ entrance}}}}
            {\|{\vec{r}_{\mu\text{ exit}} - \vec{r}_{\mu\text{ entrance}}}\|},
    \end{equation}
    \begin{equation}
    {\bar x}
        = \frac{1}{\Nfmu}\sumon{n} x_{n},
    \end{equation}
    and
    \begin{equation}
    \textsub{\sigma}{Long}
           = \sqrt{\frac{1}{\Nnmu-1}\sumon{n} \pp{x_{n} - {\bar x}}^{2}}.
    \end{equation}

\begin{figure*}
    \includegraphics[width=\columnwidth]{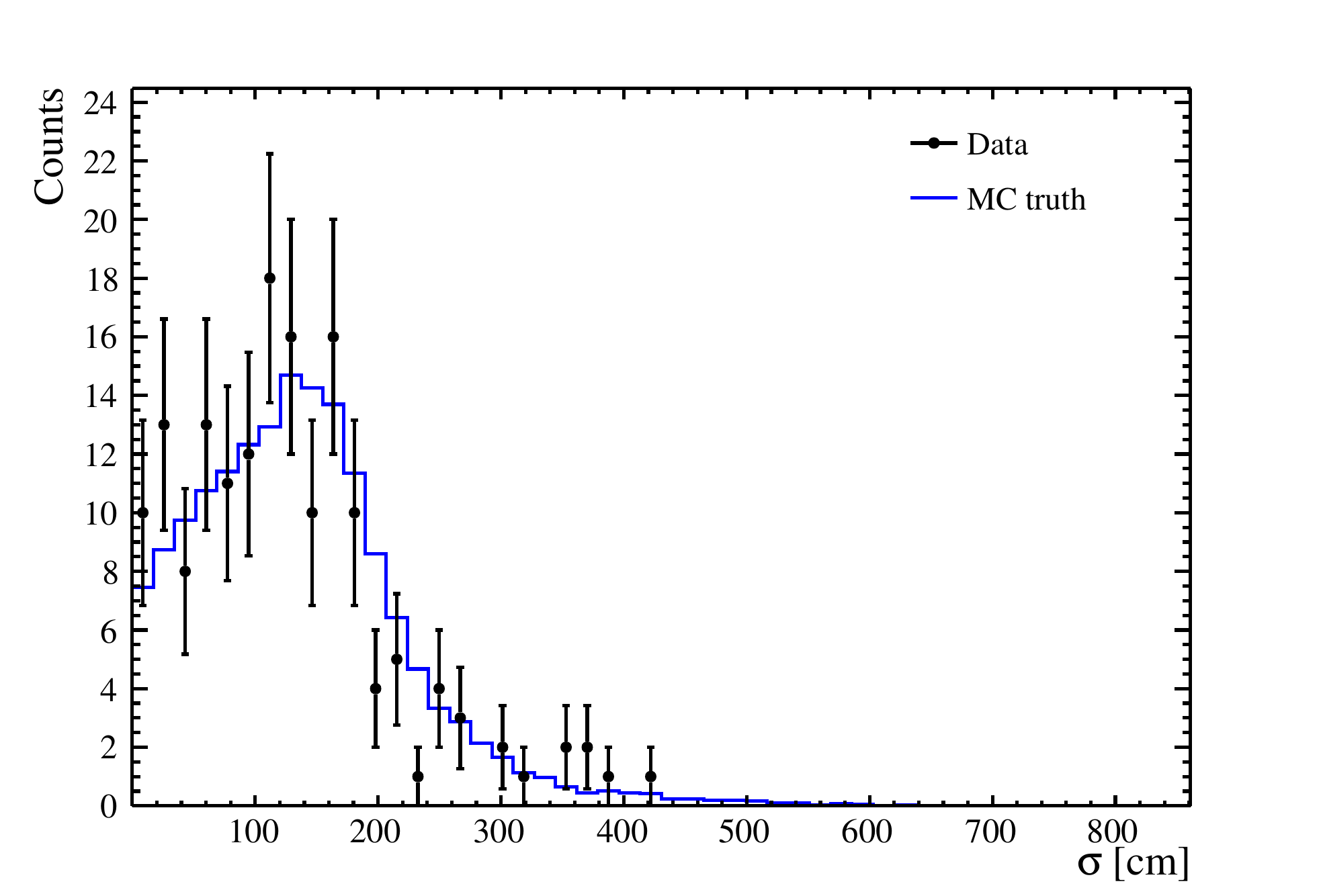}
    \includegraphics[width=\columnwidth]{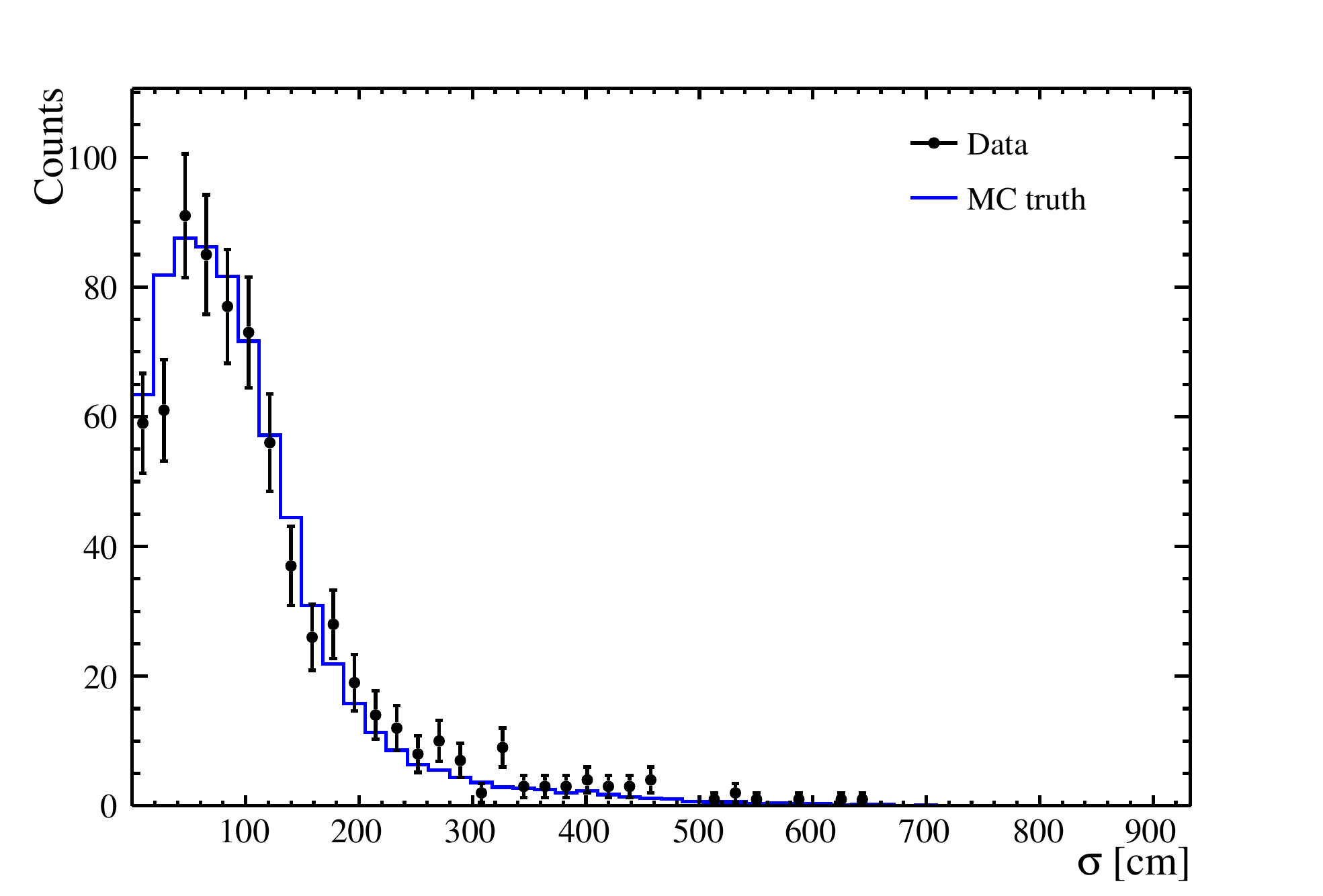}
    \includegraphics[width=\columnwidth]{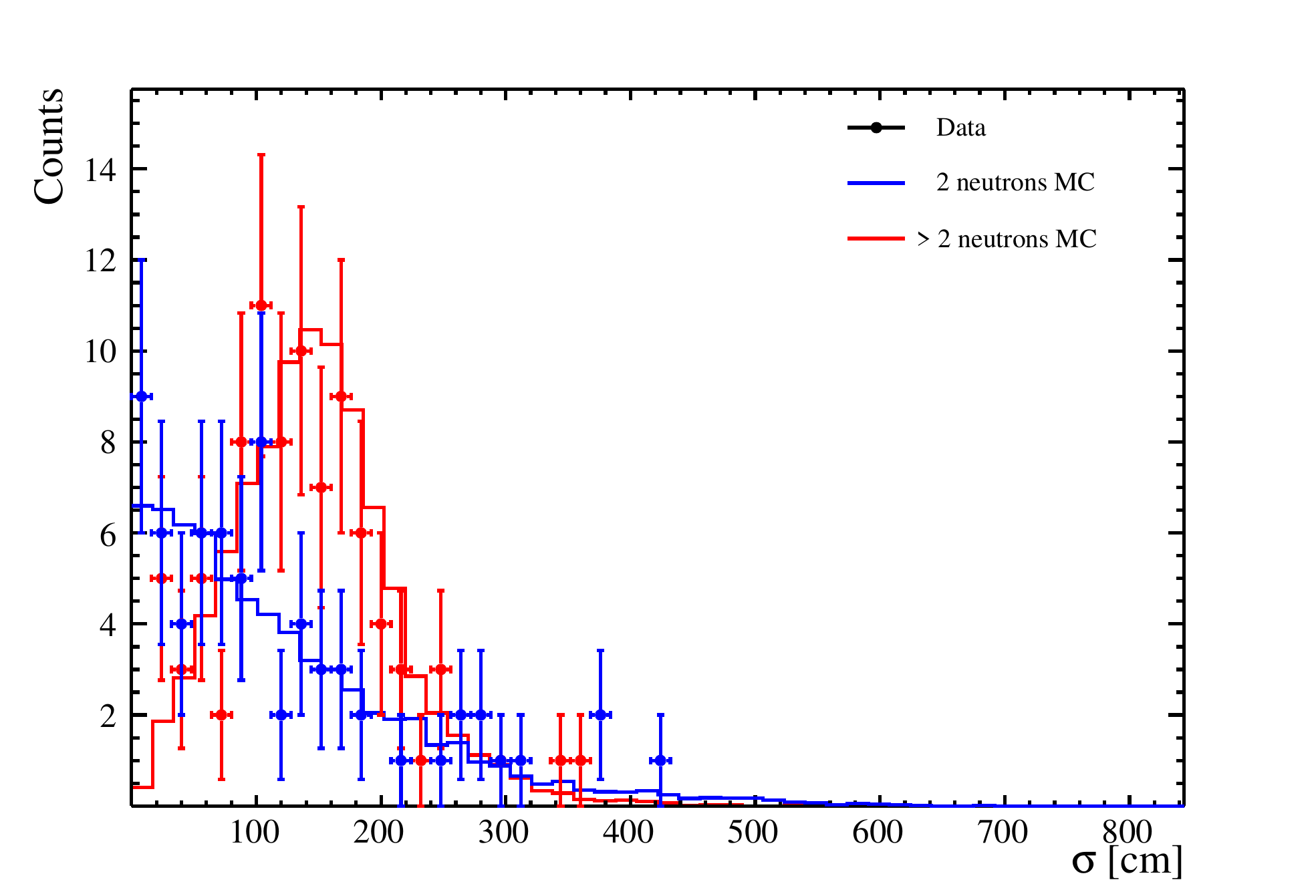}
    \includegraphics[width=\columnwidth]{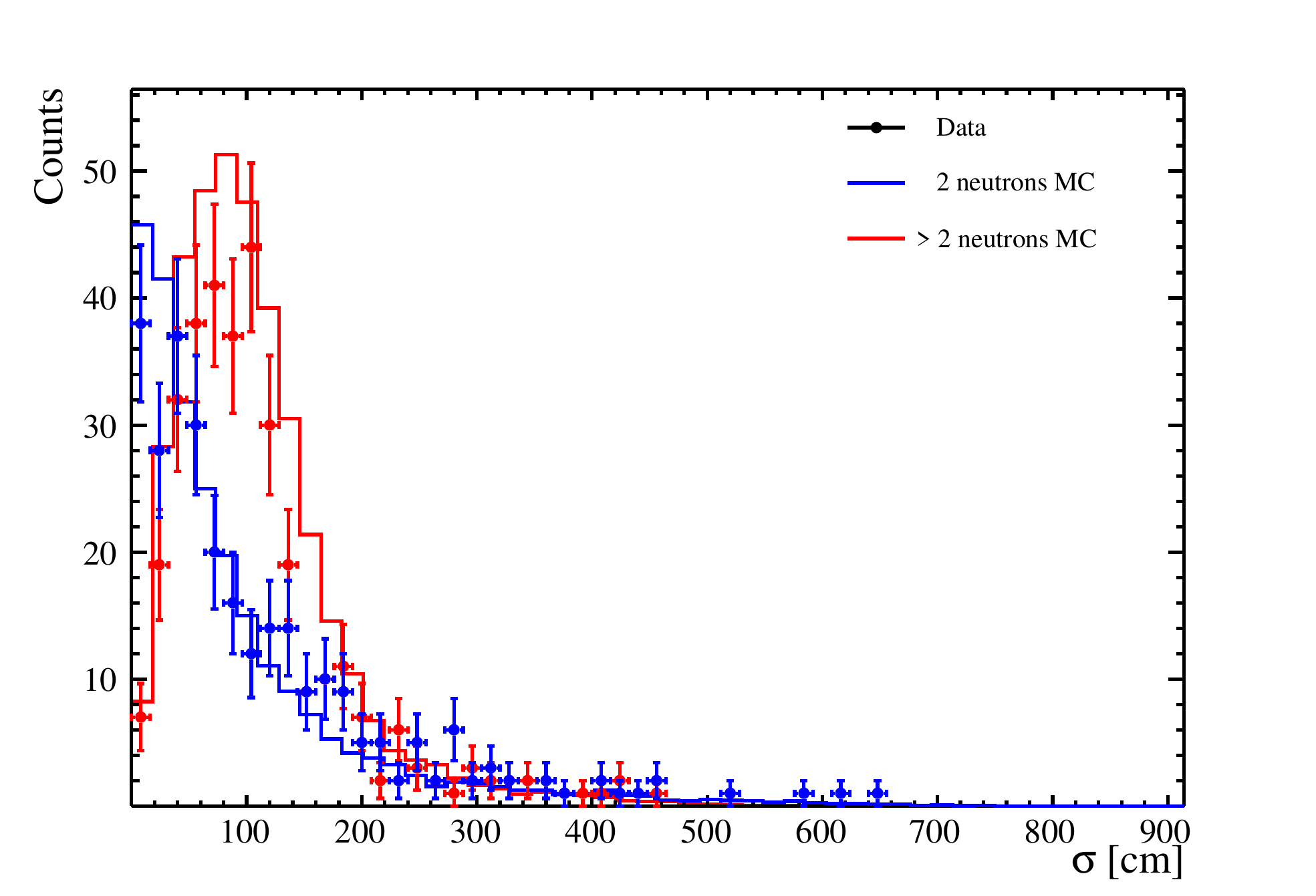}
    \caption{(Color online) Per-muon spreads of capture position measured along the track, in Phases I (left) and II (right). The bottom row shows contributions from muons of different multiplicities.}
    \label{fig:bunchinesscomparison}
\end{figure*}

    The distributions of this clustering metric in both phases are shown in \reffig{bunchinesscomparison}. The shapes of the distributions in the top panel are determined as the sum of $\chi$-distributions; a well-known result states that the variance of $n$ normally distribution samples follows a $\chi^{2}$-distribution for $n-1$ degrees of freedom. Indeed, the bottom panel of \reffig{bunchinesscomparison} shows the distributions of clustering metrics for muons broken down by multiplicity --- those followed by 2 neutrons, and those followed by greater than 2 neutrons --- and shows that the 2-neutron widths follow a falling distribution, unlike the bell-shaped curves shown for multi-neutron events.

    The mean capture profile width is $(1.28 \pm 0.06)\;\meter{}$ in Phase I, and $(1.08 \pm 0.04)\;\meter{}$ in Phase II. If interpreted as a length scale over which energy is deposited into hadronic channels, this is smaller than the expected scale for electromagnetic deposition, which in light water occurs over a range of several meters~\cite{LiBeacomShowers}.

\subsection{Lateral capture distance}
\label{sec:dperp}

\begin{figure*}
    \includegraphics[width=\columnwidth]{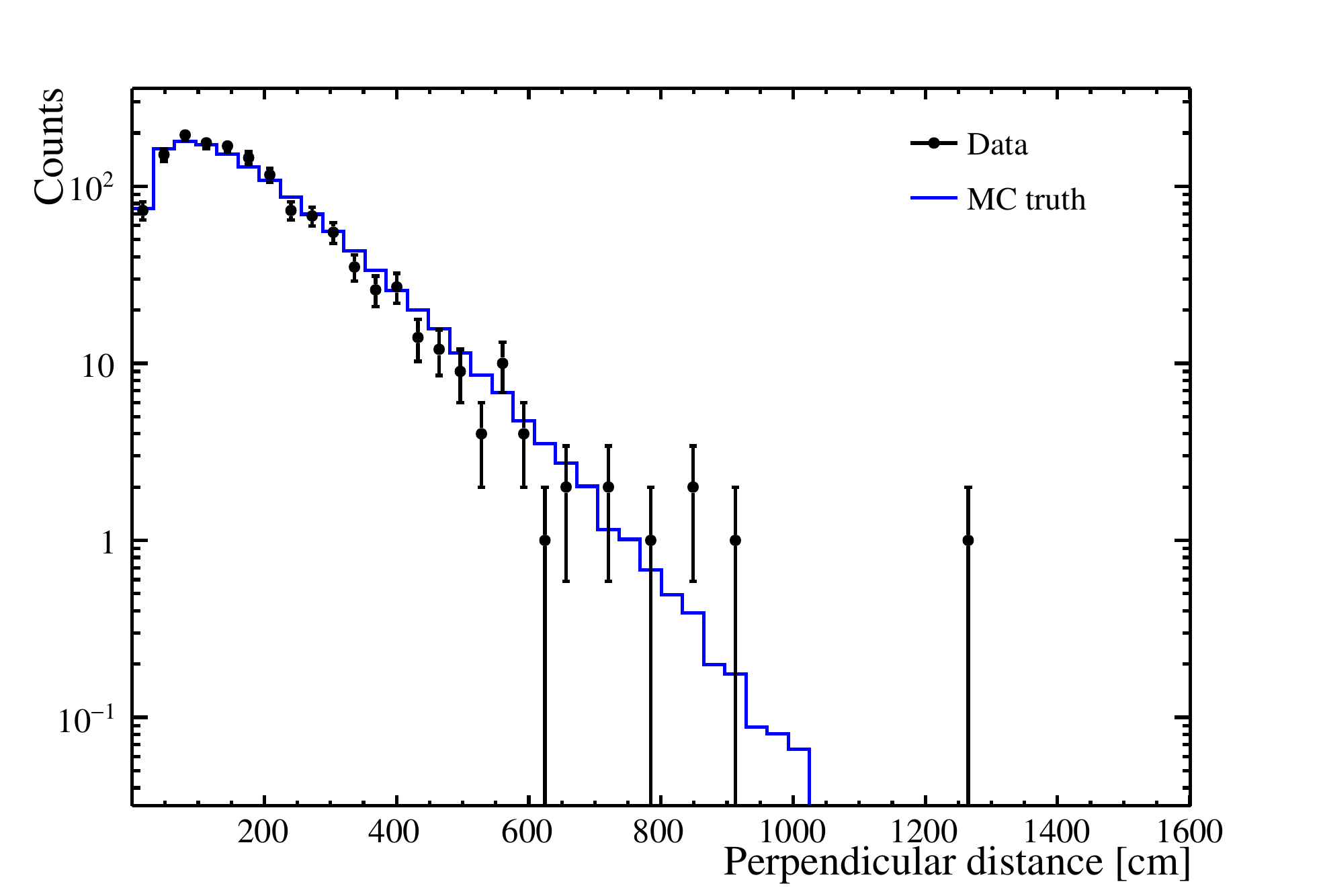}
    \includegraphics[width=\columnwidth]{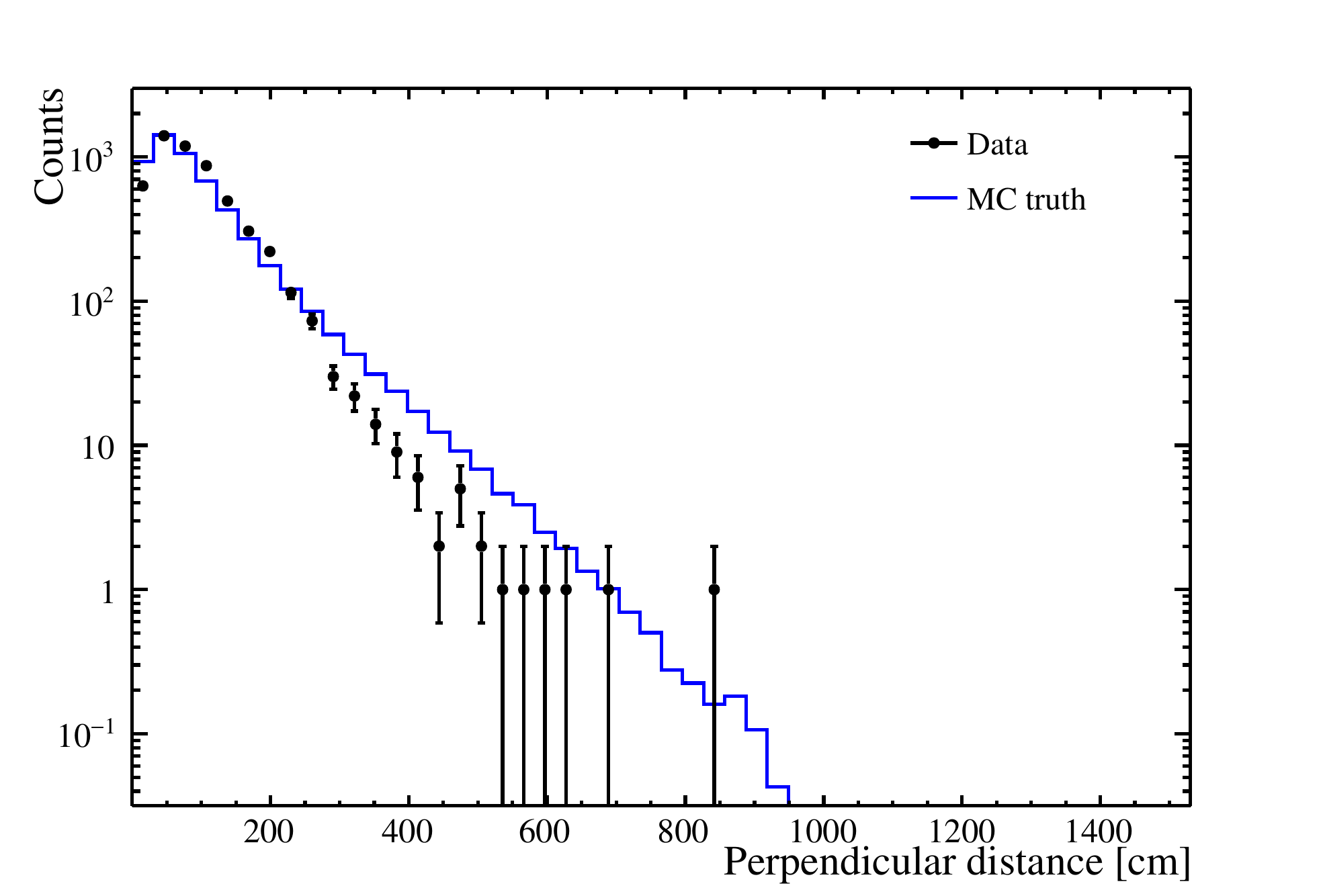}
    \caption{(Color online) Lateral capture distances from track, in Phases I (left) and II (right).}
    \label{fig:dperpcomparison}
\end{figure*}

    The distributions of the lateral capture distance from the leading track are shown in \reffig{dperpcomparison}, which follow an anticipated exponential form. The offset in exponential behavior from 0 is due both to neutrons being produced away from the track, and the distance traveled by the neutrons before thermalizing. The characteristic distances, both in data and simulation, in Phase II are reduced in comparison to Phase I, which is expected on the basis of the larger capture cross section for $^{35}$Cl than that for $^{2}$H. A single muon in Phase I preceeded a follower candidate observed more than $12\;\meter{}$ away, an extreme not predicted by the Monte Carlo. The muon did not enter the AV, and traveled only through the surrounding light water.

    The data from Phase II exhibit a rather gross difference in shape from the Monte Carlo prediction, a phenomenon not observed in Phase I. Indeed, we believe that this points to a problem with GEANT4's treatment of {\it cosmogenic} neutrons. While validations of low energy neutron transport have been performed, opportunities to benchmark models of high energy transport are scarce. It is also possible that the energy spectrum of primary neutrons determined in GEANT4 is incorrect, or that the cross sections for scattering from chlorine at high energy are not valid. No such discrepancy is observed in Phase I because low energy neutrons in deuterium experience appreciable random walks, typically several meters in length, before capturing. Any sub-meter difference in the path length traveled at high energy is masked by the effect of this relatively long random walk. Indeed, using a simple toy MC which samples high-energy transport lengths from the Phase II distributions in \reffig{dperpcomparison} and low-energy transport lengths from the distribution of random walk lengths that a neutron may experience in pure \dtwoo{}, the resulting distributions exhibit a similar level of agreement as in Phase I, in which no discrepancy is seen.

\subsection{Time delay}
\label{sec:delay}

\begin{figure*}
    \includegraphics[width=\columnwidth]{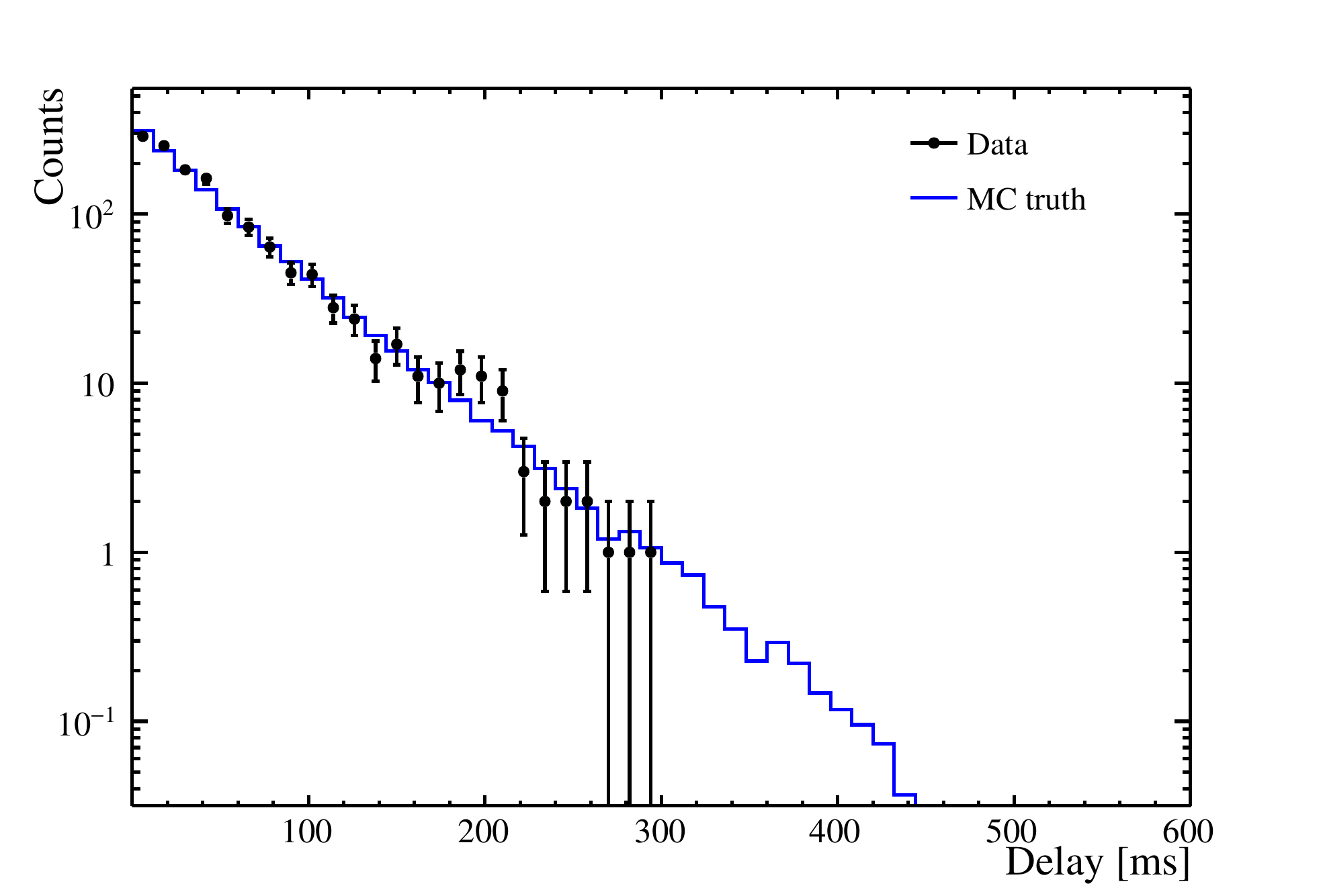}
    \includegraphics[width=\columnwidth]{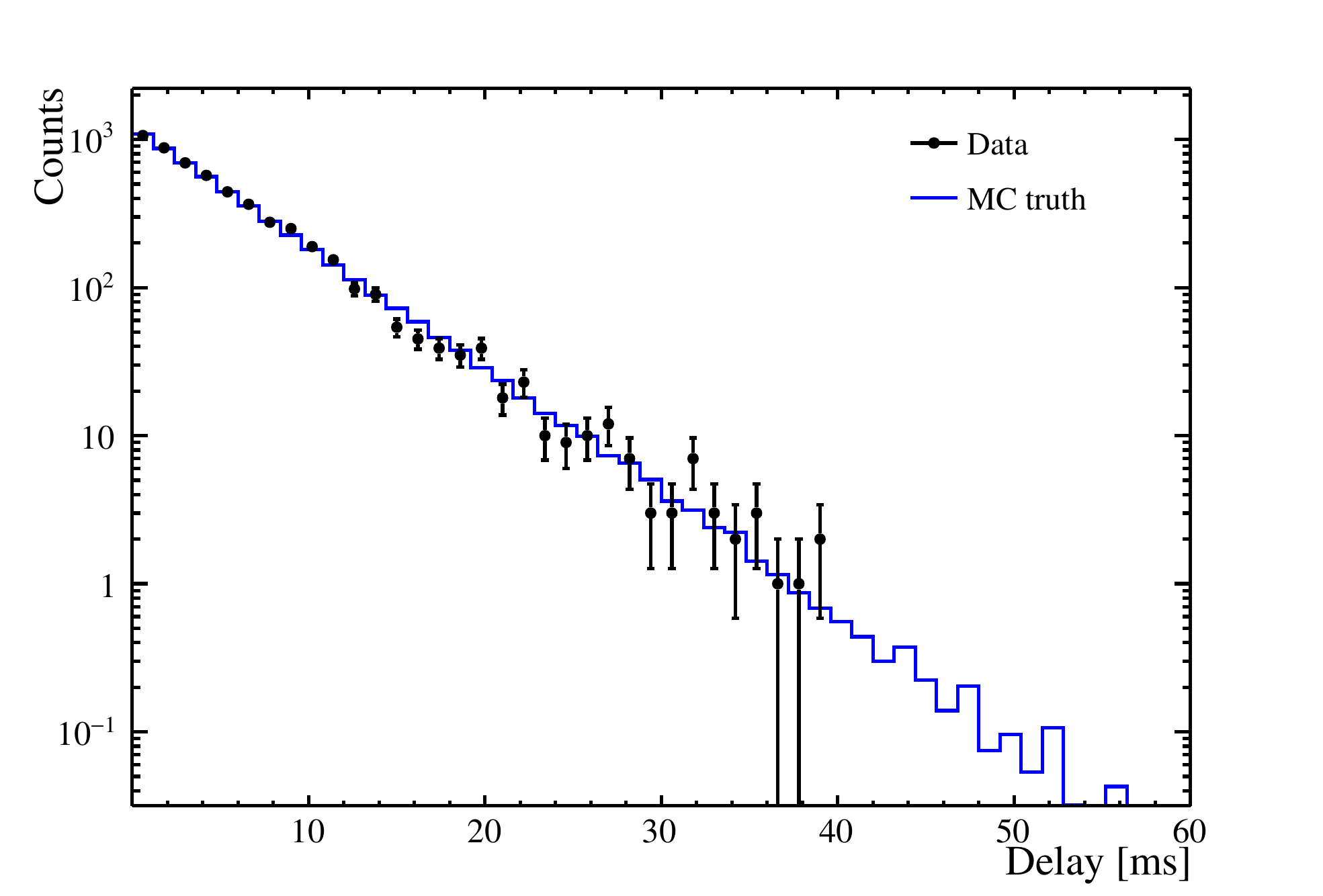}
    \caption{(Color online) Follower delay from most recent muon, in Phases I (left) and II (right).}
    \label{fig:delaycomparison}
\end{figure*}

    Distributions of the delay between a muon's passage through the detector and its follower captures are shown in \reffig{delaycomparison}. The data during each phase may be fit with a pure exponential, yielding maximum-likelihood estimators of the characteristic capture time of
        $48.5 \pm 1.3\;\milli\second{}$
    in Phase I, and
        $5.29 \pm 0.07\;\milli\second{}$
    in Phase II. While muon-induced neutrons may be
    produced with very high energies, this is in agreement with the
    previously measured capture time for $^{252}$Cf neutrons in the
    salt phase of
        $5.29 \pm 0.05\;\milli\second{}$
   ~\cite{SaltPaper}. As the thermalization time is small in comparison to the overall capture time, this agreement suggests that the modeling of low-energy neutron transport and capture are valid in the presence of chlorine, further indicating that the source of the discrepancy in lateral capture distance is in the high energy regime.

\section{Results for neutron yield}
\label{sec:yieldresults}

    The measured neutron yield values in pure heavy water and salt-loaded heavy water are found to be, in units of $10^{-4}\;\yieldunits$, $\hwtryield$ and $\saltyield$, respectively. These are to be compared with the respective values predicted by GEANT4 of $7.01 \pm 0.014\;\stat$ and $7.29 \pm 0.014\;\stat$, respectively, though it should be noted that systematic uncertainties on the simulated values may be quite large; see the extensive discussion in~\cite{KluckThesis}.

    The systematic uncertainties for this measurement are shown in \reftbl{systematics}, including uncertainties from the Monte Carlo-based capture and observation efficiencies, as well as the number of neutron-like background counts coincident with a through-going muon.

\begin{figure}
        \includegraphics[width=\columnwidth]{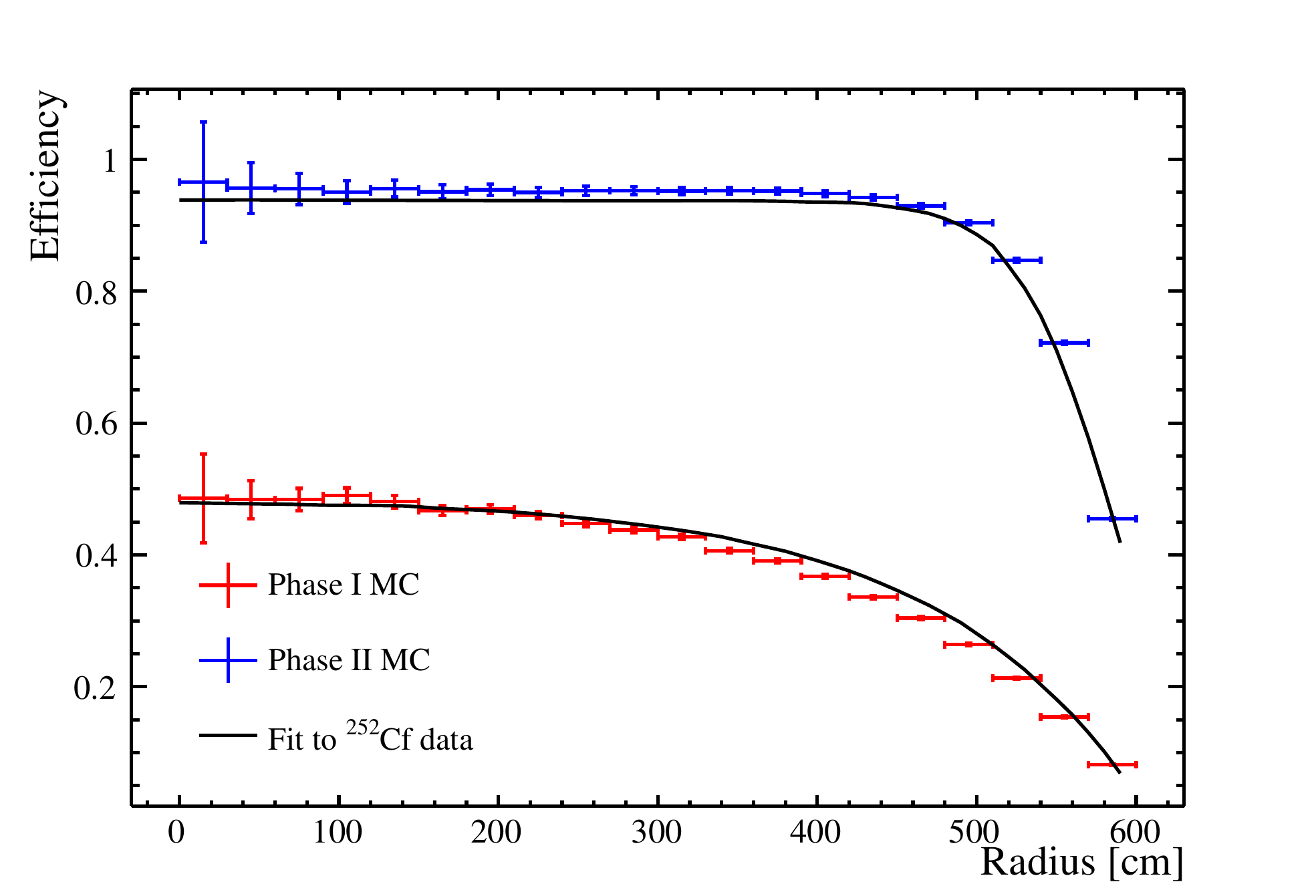}
        \caption{(Color online) Low energy capture efficiencies as calculated by simulating $^{252}$Cf-fission neutrons with GEANT4, compared with analytic fits performed to $^{252}$Cf calibration data taken during Phases I and II.}
        \label{fig:cf252}
    \end{figure}

    The dominant uncertainty is due to the Monte Carlo-based capture efficiency. A $^{252}$Cf fission source was deployed in both phases to measure a per-neutron capture efficiency for low energy ($< 15\;\MeV{}$) neutrons as a function of position in the detector~\cite{SaltPaper}. We assess an additional uncertainty on the muon-induced capture efficiency by computing a volume-weighted average of the relative error between the capture efficiency for $^{252}$Cf neutrons as reported by GEANT4 and the results of the calibration campaign, which are shown in \reffig{cf252}. While the simulation is able to reproduce the gross features of the low-energy capture efficiency in both phases, the disagreement at high radii, where the efficiency decreases substantially, causes this to be the dominant uncertainty.

    \MkTable{systematics}{l c c}{
        \- & Phase I & Phase II \\
        \hline
        Capture efficiency \
            & $\plusminus{21.7\%}{15.2\%}$ & $\plusminus{19.1\%}{13.8\%}$ \\
%       \hline
        Observation efficiency \
            & $\pm 0.4\%$ & $\pm 2.1\%$ \\
%       \hline
        Background counts \
            & $\plusminus{0.0\%}{2.4\%}$ & $\plusminus{0.0\%}{0.7\%}$ \\
        \hline\hline
        Total \
            & $\plusminus{21.7\%}{15.3\%}$ & $\plusminus{19.2\%}{14.0\%}$ \\
           
    }{Relative uncertainties on the yield measurement.}

\subsection{Evaluation of the Poisson hypothesis}

    The yield value presented above is the measurement of $Y_{n}$ (see \refeq{yield}), which is standard in the literature, and is the value appropriate when describing neutron production as a Poisson process. This can be compared to the mean per-muon yield, ${\bar Y}_{n}$ (see \refeq{simpleavg}), which in units of $10^{-4}\;\yieldunits$ is $7.62 \pm 0.89\;\stat$ and $9.32 \pm 1.22\;\stat$, in Phases I and II, respectively. The two rates are consistent in pure heavy water, but not in Phase II, where the discrepancy is $24.4\%$. The mean per-muon yield is more sensitive to high-multiplicity muons than the idealized rate, and indeed the few muons in the tail of the Phase II distribution shown in \reffig{multiplicitycomparison} are the source of this difference. Monte Carlo sampling indicates that a discrepancy this large is not unusual, and suggests that a Poisson rate, while useful for summarizing a gross production rate, should not be interpreted as a parameter fundamental to neutron production.

\subsection{Comparison to other experiments}

\begin{figure}
        \includegraphics[width=\columnwidth]{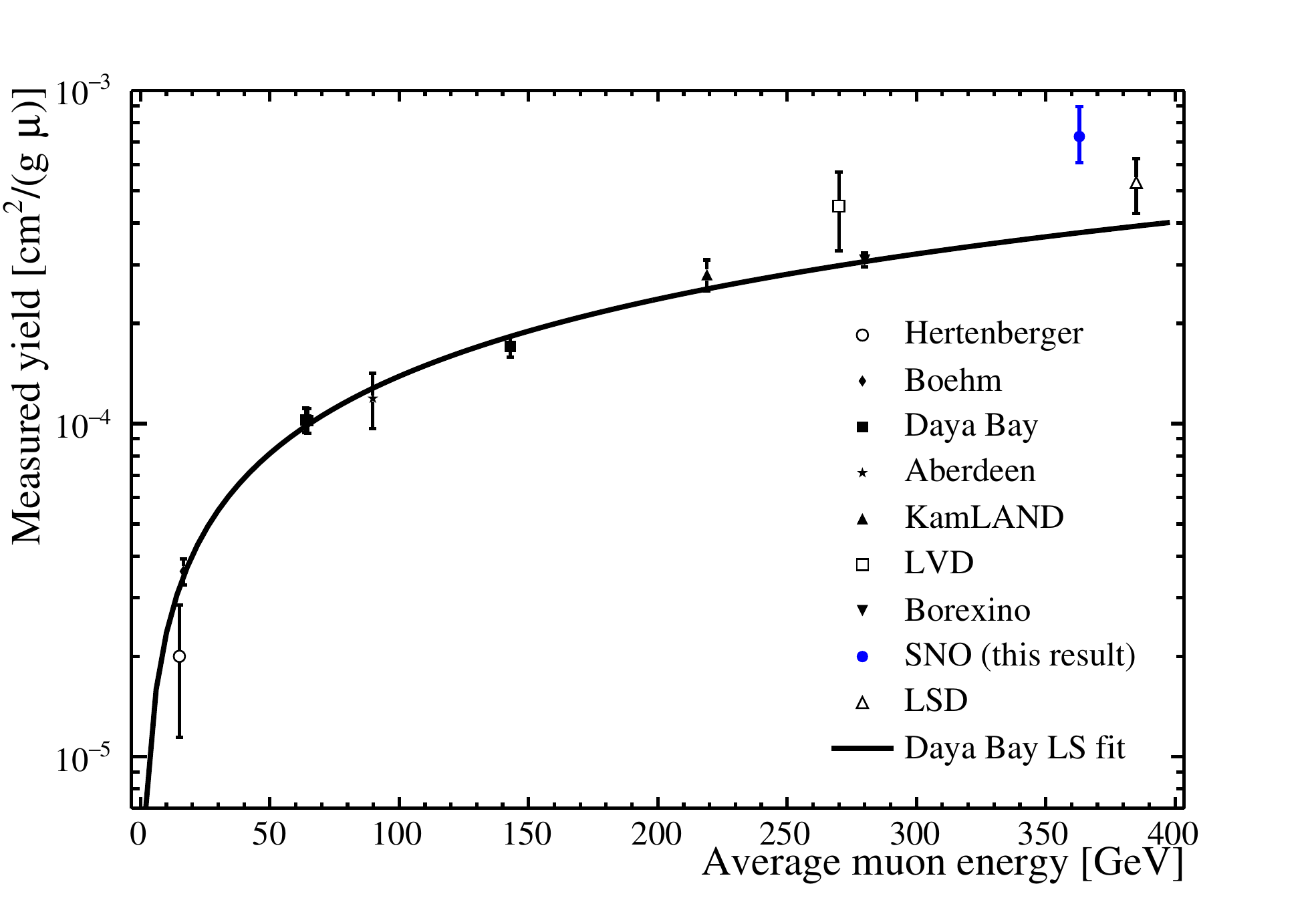}
        \caption{Power-law fit for the cosmogenic neutron yield in liquid scintillator, performed by the Daya Bay Collaboration~\cite{DayaBay}, with the SNO Phase I and LSD measurements overlaid. The SNO and LSD measurements are not included in the fit, and the target material used in SNO is different.}
        \label{fig:yieldcompare}
    \end{figure}

    While no cosmogenic neutron yield measurements have been published for heavy water, several have been performed using liquid scintillator targets. The nuclear composition of heavy water, abundant with weakly bound deuterons, differs from that of the carbon chains typically found in organic liquid scintillators, and so the results should not be compared directly. Still, the average numbers of nucleons per unit volume are comparable, and so the yields should be of similar scale. \reffig{yieldcompare} shows several yield measurements performed with liquid scintillator targets as a function of average muon energy, and a fit to a scaling law of the form $Y_{n} = a E_{\mu}^{b}$ recently performed by the Daya Bay Collaboration~\cite{DayaBay}, with both the LSD~\cite{LSD} and this measurement overlaid. The average muon energy at SNO depth was determined using the parameterization in \cite{MeiHime}. It is observed that while cosmogenic neutron production in heavy water occurs on a similar scale to the extrapolation from liquid scintillator measurements, it is enhanced, consistent with the greater average mass number. With the SNO+ experiment currently running in the original SNO cavern with plans to record data with both light water and liquid scintillator targets, it will be possible to perform additional yield measurements at this same site using multiple different materials, to further elucidate the nature of neutron production at such high energies.

\pagebreak

\section{Conclusions}

    Although the production and propagation of cosmogenic neutrons are modeled in publicly available software, such as GEANT4~\cite{GEANT4}, these models have not been exhaustively tested, particularly at the depth of SNO, due to the scarcity of experimental data. Extrapolations from more shallow experimental sites are not well understood.  SNO offers a unique opportunity to test models at this depth, and in this muon energy regime, as well as to understand this source of background events for other experiments at SNOLAB.  Community-standard simulation tools are seen to reproduce many characteristic observables of muon-induced neutrons in the SNO detector. However, some discrepancies indicate that these tools may be improved, particularly in the high energy regime. Using these simulation tools, the cosmogenic neutron yield at a depth of 5890 km.w.e. in heavy water, and heavy water loaded with 0.02\% NaCl by mass, is found to be, in units of $10^{-4}\;\yieldunits$, $\hwtryield$ and $\saltyield$, respectively.

    With many low-background experiments operating and planned in the coming decade, the measurements and model comparisons presented here are important for a better understanding of the background models used in these experiments.

%   Although the production and propagation of cosmogenic neutrons are modeled in publicly available software, such as GEANT4~\cite{GEANT4}, these models have not been exhaustively tested, particularly at the depth of SNO, due to the scarcity of experimental data. Extrapolations from more shallow experimental sites are not well understood. SNO offers a unique opportunity to test models at this depth, and in this muon energy regime, as well as to understand this source of background events for other experiments at SNOLAB. With many low-background experiments operating and planned in the coming decade, the measurements and model comparisons presented here are important for a better understanding of the background models used in these experiments.

%   Community-standard simulation tools are seen to reproduce many characteristic observables of muon-induced neutrons in the SNO detector. However, some discrepancies indicate that these tools may be improved, particularly in the high energy regime. Using these simulation tools, the cosmogenic neutron yield at a depth of $5890$~km.w.e. in heavy water, and heavy water loaded with 0.02\% NaCl by mass, is found to be, in units of $10^{-4}\;\yieldunits$, $\hwtryield$ and $\saltyield$, respectively.

\begin{acknowledgments}
This research was supported by: Canada: Natural Sciences and Engineering Research Council, Industry Canada, National Research Council, Northern Ontario Heritage Fund, Atomic Energy of Canada, Ltd., Ontario Power Generation, High Performance Computing Virtual Laboratory, Canada Foundation for Innovation, Canada Research Chairs program; US: Department of Energy Office of Nuclear Physics, National Energy Research Scientific Computing Center, Alfred P. Sloan Foundation, National Science Foundation, the Queen's Breakthrough Fund, Department of Energy National Nuclear Security Administration through the Nuclear Science and Security Consortium; UK: Science and Technology Facilities Council (formerly Particle Physics and Astronomy Research Council); Portugal: Funda\c{c}\~{a}o para a Ci\^{e}ncia e a Tecnologia. We thank the SNO technical staff for their strong contributions. We thank INCO (now Vale, Ltd.) for hosting this project in their Creighton mine.
This research used the Savio computational cluster resource provided by the Berkeley Research Computing program at the University of California, Berkeley (supported by the UC Berkeley Chancellor, Vice Chancellor for Research, and Chief Information Officer).

The authors would like to thank John Beacom and Shirley Li for useful discussions.
\end{acknowledgments}

\bibliography{cosmo}

\end{document}